# THERMODYNAMICS
## BASIC PRINCIPLES

*G. V. Skornyakov*

Ioffe Institute, Saint-Petersburg, Russia, e-mail: skorn@ioffe.ru

## ANNOTATION

Logical and mathematical aspects of the basic concepts of thermodynamics are considered.

## FOREWORD

This treatise is in no way meant to serve as a help in gaining acquaintance with the theory of thermal processes. Discussion of the fundamentals of a theory can bear fruit only if led with competent enough people. The author starts with a premise that the Reader embarking on a study of this book has a general knowledge of this theory within the program of the course of General physics offered at a Physical Faculty of a University [1]. The major motivation that has prompted the Author to write this book was the inconsistency of the second law of thermodynamics as a law of Nature [2].

The logical structure of thermodynamics is often compared to that of geometry in the consistent and strict way it emerges from the two postulates (laws of thermodynamics). The conclusions drawn in thermodynamics found convincing support in the truly immense variety of observations and deductions, so that one could hardly find today a physicist, chemist or biologist who would question the validity of thermodynamics.

Rather than being only a subject of scientific studies, thermal processes truly pervade all aspects of our life. Thermodynamics is generally considered to be a scientific basis underpinning heat-power engineering, and conversion of heat to work, the main problem facing thermodynamics. Conversion of heat to work by various heat engines had been enjoying worldwide industrial application long before thermodynamics was formulated. From the very beginning, the major motivation underlying thermodynamics had been a search for the most efficient methods that could be applied to this conversion. But it is thermodynamics that grew to become an insurmountable monolithic wall separating the truly inexhaustible ocean of thermal energy surrounding us from the possibility of its use in technology.

Abandoning the second postulate as a universal law of Nature would at first glance seem to imply a revision of all already well-established conceptions about the world surrounding us. In actual fact, however, the situation is not that tragic. Indeed, the overwhelming majority (if not all) of the processes converting heat to work which were considered in thermodynamics relate to single-parameter systems. For these processes, the second law certainly does hold. But if the number of external parameters of a system is larger than one, the applicability of the second law to it requires a rigorous proof. Said otherwise, the second law stated in its universally accepted form is nothing else than extension of the commonplace to the domain of Unknown.

The second law is a deep but regrettably not the only delusion in the present concepts bearing on the nature of thermal processes. Another one is actual identification of the distribution function of a nonequilibrium system with that of probabilities.

It is a commonplace belief that the "physical essence" of the fundamental concepts of thermodynamics is revealed by statistical physics. The purpose of the molecular kinetics theory lies not only in explaining the concepts making up the framework of the phenomenological theory which had been thermodynamics from the very beginning, but in defining the limits of their applicability.

Thermodynamics does not formulate the equations governing the evolution of the state of a thermodynamic system in time. It only specifies the general principles this evolution has to obey in order to progress. This characteristic feature of thermodynamics generated a certain feeling of dissatisfaction in many researchers. A number of concepts well established in thermodynamics (such as temperature, entropy) penetrated into the description of notoriously nonequilibrium processes. What is more, even formulation of the basic laws of hydrodynamics turned out to be impossible without their use. Many concepts that originally had come from thermodynamics have so deeply penetrated into our consciousness as to become frequently used, so to say, by default.

It is an analysis of the fundamentals of thermodynamics that form the basement of the categories it involves that the present treatise is devoted to. Particular emphasis is placed on the mathematics underlying the concepts employed in thermodynamics. No specific problems are considered.

# INTRODUCTION

The beginning of the XX century was marked by two events which changed radically the overall scientific vision of the material world and of the nature of space and time, more specifically, by formulation of quantum theory and development of the theory of relativity. The theory of relativity joined the space and the time, removed the well-known contradiction between classical mechanics and electrodynamics, and, finally, after the general theory of relativity had been developed, gave birth to the present ideas connecting matter, space and time which form the basis of our knowledge of the structure and evolution of the Universe. Still more radical changes into the character of the overall scientific thinking were introduced by quantum theory which molded a microworld driven by interrelations of a qualitatively different character from those that became familiar in the many years of studies of the nature of the macroworld.

These two greatest revolutions in the history of natural science revealed, however, a common characteristic feature in that they, rather than introducing noticeable changes into the laws established before, only imposed rigorous bounds on the area of their applicability. More than that, the laws of classical physics turned out to be not more than a limiting case of new laws which are valid for changes in velocity small compared to the velocity of light and changes in action large compared to the Planck constant. Considered in the context of the correspondence principle, this relation between the old and the new theories is the necessary criterion of the objective character of knowledge and of the everlasting value of science.

Thermodynamics is a major section of classical physics. The origin of thermodynamics as science is dated back to the beginning of the XIX century. It would not be an overstatement to say that the groundwork for the fundamentals of thermodynamics was laid by Sadi Carnot; indeed he was the first to formulate the basic postulate which was since called the Second Law of thermodynamics. The First Law of thermodynamics was formulated much later, in the mid-XIX century. Thermodynamics was borne and continues to exist as a phenomenological theory resting directly upon experimental data, because it was essentially their theoretical generalization on which it was built. The molecular kinetics theory provided a theoretical basis for the main phenomenological postulates of thermodynamics. Creation of quantum theory completed the construction of the harmonious building of thermodynamics which rested on the reliable basement of experimental data.

Thermodynamics has the reputation of a theory built around the strictest possible logical structure. The logics underlying it is often compared to that of geometry. Like geometry, it rests upon an axiomatic basis. In the width of the scope it encompasses thermodynamics has no equals among physical theories. A. Einstein wrote in this connection [3]:

"A theory is the more impressive the greater the simplicity of its premises is, the more different kinds of things it relates, and the more extended is its area of applicability. Therefore the deep impression which classical thermodynamics made upon me. It is the only physical theory of universal content concerning which I am convinced that, within the framework of the applicability of its basic concepts, it will never be overthrown (for the special attention of those who are skeptics on principle)."

Now what are the most remarkable features A. Einstein singled out in thermodynamics? Recalling the creation of the theory of relativity he stressed [3]:

"…Only the discovery of a universal formal principle could lead us to assured results. The example I saw before me was thermodynamics. The general principle was there given in the theorem: the laws of nature are such that it is impossible to construct a perpetuum mobile (of the first and second kind)… The universal principle of the special theory of relativity is contained in the postulate: The laws of physics are invariant with respect to the Lorentz transformations… This is a restricting principle for natural laws, comparable to the restricting principle of the non-existence of the perpetuum mobile which underlies thermodynamics."

It is thermodynamics, or, to be more exact, thermodynamics of radiation that formed the first bridge connecting the classical with quantum physics. Only the firm conviction that thermodynamics is absolutely valid could form a firm basement for a radical transformation of the fundamental physical ideas, a transformation that was required by quantum theory from the very beginning. The triumph of quantum theory only strengthened this confidence.

On the other hand, one could not ignore indications that strict adherence to a viewpoint based on the laws of thermodynamics can result in a striking contradiction with the pattern of the world as we see it, and even with the fact itself of our existence. It appears only natural to ask could not these contradictions be the consequence of our using the laws of thermodynamics beyond the boundaries of their applicability?

Thermodynamics occupies by right a particular place in the realm of physical disciplines. The character itself of thermodynamic description differs radically from that of other physical theories, such as mechanics or electrodynamics (to the extent to which they allow neglect of thermal processes). A physical theory rests ordinarily on the concept of the state of the object under study. The subject of the theory lies in identifying the state and studying the process of its variation with time. In thermodynamics, the question is posed in an essentially different way.
Thermodynamics studies macroscopic bodies composed of an enormous number of atoms and molecules, but describes their state using a very limited number of independent variables accessible to an experimental study. In the case of gases, there are only two such variables. Gases represent a convenient example of a graphic application of the main principles underlying thermodynamics, which is bound tightly to statistics. The concept itself of an object in thermodynamics is inseparable from the conditions in which it exists.

Thermodynamics combines tightly intertwined physical quantities differing in nature. An important part is played by external parameters whose variation is directly connected with the work done by the system (or on it) [1]. In the case of gases in a closed vessel, such an external parameter is volume. Another important characteristic is internal parameters which define interaction of a system with the surroundings. For gases, the internal parameter is pressure. Both the volume and the pressure of a gas are determined by external conditions. Moreover, it is these conditions that specify fully the assignment of the parameters of a system to the external or internal domain. The processes at work in a gas can be studied both at a constant volume and a constant pressure. In the latter case, volume becomes the internal, and pressure, the external parameter. This is why it is better to discuss thermodynamic parameters of a system without refining which of them are external, and which, internal [1].

The variables defining the thermodynamic state of a system are essentially characteristics of a complex statistical system averaged over a more or less extended interval of time and, therefore, are subject to fluctuations rather than being set precisely. Close to critical points, the fluctuations may reach macroscopic levels, thus making the principal premises underlying thermodynamics, with the exception of the energy conservation law, inapplicable.

Far from all macroscopic processes are truly thermodynamic. The processes at work in statistically nonequilibrium systems are notoriously not among them. At the same time a major concept underlying thermodynamics is that of thermal equilibrium, by which at fixed external conditions an isolated system relaxes eventually, irrespective of its initial state and at fixed external conditions, to a state that does not change thereafter. This statement is termed sometimes the zeroth law of thermodynamics.

The major topic of thermodynamics is essentially investigation of the behavior of macroscopic bodies in the limiting case of a slow variation of the variables defining the bodies' state. Such quasistatic processes are reversible [1]. But this means that what one is used to call thermodynamics is in actual fact thermostatics. In the domain of thermostatics, there can be no heat and material flows.

## THERMODYNAMICS AND STATISTICS

Introduction of parameters defining the state of a thermodynamic system relates thermodynamics only outwardly to other physical theories. By specifying the parameters of a system, one fixes not one but rather a set of states corresponding to the parameter set specified. Rather than defining the state chosen out of this set, thermodynamic parameters identify only the sampling space [4] to which they belong. In the particular case of gases, a state is described by a point in the phase space of the system. Only macroscopic parameters remain invariable in thermodynamic equilibrium. As for the phase trajectory in the space of microscopic variables, its specific form does not in any way affect the macroscopic parameters.

Motion of gas molecules obeys the laws of classical mechanics. They define a dynamic system with an integral invariant (the Liouville theorem [1]). The extremely complex, entangled pattern of the motion of molecules suggests resorting to the theory of probability for its description. This appears all the more justified that classical mechanics is capable of only an approximate description of molecular motion. A consistent quantum-mechanical analysis [5] not only correlates well with the classical approach but makes it possible to eliminate its inherent contradictions (the Gibbs paradox). One should, however, bear in mind that the theory of probability is by no means an all-embracing theory of stochastic systems and processes. Not any random quantity can be characterized by a definite probability. We are deferring a more comprehensive analysis of this point to the Conclusion.

For given parameters of a system, complete statistical equilibrium requires a considerable time for it to set in. This time should be long enough for all molecules to interact efficiently with one another. If thermodynamic categories were applicable only to such systems (strictly speaking, this is exactly the case), thermodynamics would not be enjoying presently the significance it has acquired in the nearly two centuries of its existence. In actual fact, however, practically all molecules interact efficiently only with those located at a distance of the order of a few mean free path lengths. In a macroscopic volume with linear dimensions on the order of several mean free path lengths, a close to complete equilibrium sets in in about a few mean free path times. Such systems are commonly considered to be in local equilibrium. And it is the existence of local equilibrium that makes it possible to extend the thermodynamic approach to more or less nonequilibrium systems, while at the same time setting a firm boundary to attempts of considering the laws of thermodynamics to be absolute concepts.

# SINGLE-PARAMETER SYSTEMS

The characteristics of a system are governed not only by the material of which it is made but by the way in which it is separated from the surroundings. This "partition" may have a variety of specific properties. If the system is placed into a fixed heat-impermeable, adiabatic vessel, the state of the surroundings does not affect in any way the processes at work in it. If the walls of the vessel do conduct heat, the temperature of the system at equilibrium will be equal to that of the surroundings. Also, if the elements of the "partition" are capable of moving freely, the pressure in the system will coincide with that of the surroundings. The latter condition is particularly essential for gases. A perfect gas offers us a convenient possibility of illustrating graphically all the main thermodynamic categories.

One of the major thermodynamic categories is the amount of heat $Q$. The first law of thermodynamics relates in a well known way a small change in the amount of heat in a thermodynamic system to that of its internal energy and the work done on it (or by it). If we choose volume as the external parameter, this relation can be written as

$$\delta Q = dE + PdV, \qquad (1)$$

where $\delta Q$ is the amount of heat transferred to the gas, $P$ is the pressure of the gas, and $dV$ is the change of its volume. The variation of the amount of heat is not, however, a differential of any function. And this is what accounts for the possibility itself of the conversion of heat to work in a cyclic process by means of a heater and a cooler. If a system is described by one external parameter only, a small change in the amount of heat, while not being a total differential, has nevertheless an integrating factor (divisor). It is the non-negative function of internal energy and of an external parameter, which is the integrating divisor of the Pfaff form (1), that is usually referred as the absolute temperature.

The energy and temperature of a gas are closely related

$$dE = C_V dT, \qquad (2)$$

where $C_V$ is the heat capacity at constant volume. $C_V = Nc_V$, where $N$ is the number of molecules, and $c_V$ is the heat capacity per molecule. Thus the law of energy conservation can be rewritten in the form

$$\delta Q = C_V dT + PdV. \qquad (3)$$

Here temperature and volume are independent variables. For independent variables one could choose temperature and pressure as well. Recalling the equation of state for an ideal gas, $PV = NT$ (the Boltzmann constant is unity), the first law can be cast in the form

$$\delta Q = C_P dT - VdP, \qquad (4)$$

where $C_P$ is the heat capacity at constant pressure. $C_P = Nc_p$, $c_p = c_V + 1$. The heat capacity at constant pressure is larger by far than that at constant volume, because, in accordance with the equation of state for a gas, its volume increases under heating, and the system does work on the outer medium, which requires expenditure of energy.

The first law can be written also in the form
$$\delta Q = TdS, \qquad (5)$$
where $S$ is thermodynamic entropy, i.e., the function whose differential is equal to the right-hand part of Eqs. (1) or (3) divided by temperature. It should be borne in mind that there is no general method by which one could find the integrating factor, and that in the cases where it does exist, it is defined to within a factor equal to an arbitrary function of entropy. This implies that entropy also cannot be uniquely determined. The only thing of importance here is the fact itself of their existence. Despite the ambiguity of its definition, the entropy of an equilibrium system, irrespective of the way in which it was determined, may serve as an independent variable, along with the volume, pressure and temperature. No matter how the temperature was found, entropy can be determined only to within an arbitrary constant. As for temperature, for its unique determination even one instrument whose readings depend on temperature will be enough. In thermodynamic equilibrium, the temperatures of all parts of a body are the same. By bringing it in thermal contact with the object under study and waiting long enough for the equilibrium between them to obtain, one can determine readily the temperature of the object. A gas thermometer can be such an instrument.

The ambiguity about the definition of temperature offers a possibility of using any scale of its measurement that has been traditionally employed. In what follows, we are going to understand by the term "temperature" the absolute temperature in degrees Kelvin. Although the energy and the temperature are intimately connected, these concepts are essentially different. Energy is an additive quantity, whereas temperature is transitive.

It is the temperature rather than internal energy that is convenient to employ as one of the principal independent variables in thermodynamics. Energy does not have an unambiguous definition at all (if we disregard here the theory of relativity) and, just as entropy, does not belong to the realm of directly measurable quantities. Only a change in energy can be recorded by some means. In contrast to entropy, each molecule has an energy, whereas entropy, similar to temperature, is a characteristic of a macroscopic object.

Fairly small (but macroscopic) bodies reach equilibrium in a comparatively short time. Thus parts of a body that is far from equilibrium may reside in states of local equilibrium. For a fixed external parameter, the temperatures of all parts of a nonequilibrium systems may, generally speaking, vary in the course of relaxation. It is commonly assumed that at equilibrium the entropy reaches a maximum. There are, however, no grounds for such a statement. Entropy was defined for a system at equilibrium. In order for a function to have a maximum at a point, it has to be defined in its vicinity. The properties of the whole are not determined by any set of properties of its parts while being dependent on them. That the entropy of an equilibrium system is, besides energy and volume, an additive quantity and equal to the sum of the entropies of its constituent parts does in no way imply that the entropy of a nonequilibrium system can be defined as the sum of entropies of locally equilibrium parts of the system. Indeed, we do not feel surprised by the fact that a nonequilibrium system cannot be characterized by one temperature. Neither does it have entropy.

## MANY-PARAMETER SYSTEMS

The statistical approach is applicable both to a study of equilibrium states of macroscopic objects and quasi-static processes and to an analysis of any states and processes. In the particular case of gases, the subject for a study here is the distribution function of molecules in a volume and in the momentum space. The variation of the distribution function in time is specified by a kinetic equation. At equilibrium, the distribution function does not vary despite the incessant motion of molecules. The distribution function and the probability for a molecule to reside in an element of the phase volume coincide (more exactly, they are proportional). In any other state of the system, the distribution function has nothing in common with concepts of the theory of probability.

The most important concept in the theory of probability is entropy, which is defined as the mean value of the log probability [4]. This suggests immediately the possibility of definition of entropy through the mean value of the logarithm of the distribution function of a generally nonequilibrium but locally equilibrium system. As the system approaches equilibrium, the entropy defined in this way does indeed grow, provided the internal energy is constant (the H-theorem of Boltzmann). At the same time the probabilities relating to an equilibrium system do not coincide with those of parts of a nonequilibrium system. They have an *a posteriori* rather than *a priori* character and depend on the values of macroscopic parameters characterizing the parts of the system, which are related to the parameters of its other parts. The interaction of macroscopic parts of a nonequilibrium system is governed in the case of gases by the laws of hydrodynamics rather than by those of the theory of probability. Breaking up into parts assumes usually by default assigning a number of molecules to a part. But molecules are not carriers of any thermodynamic characteristics. Without a comprehensive definition of the way by which a system is to be broken up into parts there is hardly any sense in discussing parts of the system.

Let us illustrate this in a number of simple examples. Consider a horizontally arranged cylinder with three pistons, the right- and left-hand ones being actuating, and a third piston that separates them. Both parts of the volume contained between the actuating pistons are filled with gases. The gases in the compartments may be either identical or different. If the separating piston is fixed and adiabatic, the gases in both parts of the cylinder make up two totally independent thermodynamic systems. Now if the separating piston is fixed but conducts heat, or can move and is adiabatic, both gases form a common system. In the first case, both gases have the same temperature in equilibrium, while in the second, their pressures are equal. Accordingly, the pressures and temperatures in the two parts can be different. In the case of a movable and heat conducting separating piston, both the temperatures and the pressures in both parts are the same.

In the absence of a separating piston, or if it is movable and heat conducting, we have a single-parameter system. In all the other above cases, the equilibrium conditions have to be identified, generally speaking, with different values of the transitive variables. The system becomes a two-parameter one. The external parameters are now the positions of the left- and right-hand actuating pistons. In the case of a fixed heat-conducting separating piston, the two parts of the system have the same temperature, while the pressures in them may differ. If the piston is movable and adiabatic, the pressures are the same, but the temperatures may be different. As thermodynamic variables, it appears appropriate to choose in the first case the common temperature and both volumes, and in the second, the common pressure and both temperatures. One can readily see that because of the heat capacity of a gas being independent of volume, in the first case the Pfaff integrating factor always exists, while in the second, it exists only if the two gases have equal heat capacities. The absence of an integrating factor translates into the absence of entropy. An attempt of "redefining", so to say, the entropy of a system as a sum of the entropies of its constituting parts [6] makes no sense.

The thermodynamic entropy and the entropy dealt with in the theory of probability are essentially different concepts. In this treatise the term "entropy" assumes, unless otherwise specified, the thermodynamic entropy defined with a certain degree of arbitrariness. Now as for the entropy in the probability theory, it is defined rigorously and unambiguously and covers any situations obeying the laws of the probability theory.

Thus, adding just one parameter permits one to produce a system which has at equilibrium no common temperature and, thus, no entropy. It would appear that the absence of entropy is caused by thermal inhomogeneities. In thermally homogeneous systems, however, there may likewise be no integrating divisor of the Pfaff form, and, hence, no entropy altogether. Nevertheless, being a measurable quantity, the temperature of the system certainly does exist irrespective of whether the Pfaff form has an integrating divisor or not.

Systems with an arbitrary number of external parameters have the Pfaff form

$$\delta Q = dE + \sum_{i=1}^{k} A_i(a_1, a_2, \ldots a_k, E) da_i \qquad (6)$$

where $\delta Q$ is the amount of heat added to the system, $dE$ is the change of its internal energy, $da_i$ are the changes of the external parameters of system, $A_i$ are the corresponding generalized forces, and $k$ is the number of independent external parameters [1].

In the general case of many-parameter, thermally homogeneous thermodynamic systems the integrating divisor of the Pfaff form (6) does not exist at all. For $k > 1$, the form (6) has an integrating divisor only if the Frobenius conditions [4], which impose fairly rigorous limitations on the $A_i$ functions, are satisfied. For an arbitrary Pfaff form

$$\omega_n = \sum_{i}^{n} y_i(x_1, x_2, \ldots x_n) dx_i \qquad (7)$$

where $n$ is the number of independent variables, the Frobenius conditions assume the form [4, 6]

$$y_\alpha \left( \frac{\partial y_\gamma}{\partial x_\beta} - \frac{\partial y_\beta}{\partial x_\gamma} \right) + y_\beta \left( \frac{\partial y_\alpha}{\partial x_\gamma} - \frac{\partial y_\gamma}{\partial x_\alpha} \right) + y_\gamma \left( \frac{\partial y_\beta}{\partial x_\alpha} - \frac{\partial y_\alpha}{\partial x_\beta} \right) = 0 \qquad (8)$$

$$\alpha, \beta, \gamma = 1, 2, \ldots n \qquad \alpha \neq \beta \neq \gamma \neq \alpha$$

The Frobenius conditions for the $y_i$ coefficients of the Pfaff form (7) break up into a set $n!/[(n-3)!3!] = [n(n-1)(n-2)]/6$ of conditions imposed on each triple of the coefficients. While the possibility of satisfying with a set of $n$ functions the number of conditions much larger than $n$ cannot be excluded, it appears rather problematic.

## THE SECOND LAW

There are more than a dozen different equivalent formulations of the second law of thermodynamics. As an argument in its favor one even raises the decision made by the French Academy of Sciences in the XVIII century, long before the thermodynamics itself came into existence. This is often accompanied by references to experimental data. The statements of R. Clausius and W. Thomson concerning the Second Law of Thermodynamics [7], while at first glance being based on experimental data, in actual fact draw from a deep conviction of the impossibility of building a perpetuum mobile of second kind, whose construction at that time could not even be considered in the least bit seriously. Accordingly, there are no grounds for the statement of E. Fermi [8]: "The experimental proof of the validity of the second law consists primarily in all attempts at building a perpetuum mobile of second kind having met with failure". Even if such attempts had somewhere been undertaken (and of this we do not have reliable information), their failure can in no way be considered as supporting such a far-reaching conclusion.

The concept of entropy introduced by R. Clausius, which is intimately associated with the second law, resembles the entropy featuring in the theory of probability only by perception. It does not have the extent of generality in description of natural phenomena the thermodynamics claims. C. Truesdell wrote in this connection [9]:

"Seven times in the past thirty years have a tried to follow the argument Clausius offers to conclude that the integrating factor T exists *in general*, is a function of temperature alone, and is the same for all bodies, and seven times has it blanked and graveled me… I cannot even explain what I cannot understand."

The character itself of formulating the theory of thermal processes, which originates in the works of the founders of thermodynamics and is still persisting up to the present days, gave grounds to C. Truesdell to call thermodynamics "the dismal swamp of obscurity" [9]. In his Nobel Prize lecture, I. Prigogine, while not casting doubt on the laws of thermodynamics as such, stressed that "150 years after the second law had been formulated, it still resembles a program rather than a rigorously defined theory in the sense one usually understands this concept" [10].

Treating the laws of thermodynamics as a generalization of experimental data, an approach that has gained a wide popularity, does not enjoy a serious enough base. Suffice it to say that in the mid-XIX century the relative accuracy with which the mechanical equivalent of heat could be determined amounted to tens of per cent [11]. In this situation, regarding even the first law as an experimental fact would be premature, to say the least. In essence, however, the first law is a direct consequence of the physical laws being time independent. Therefore, the accuracy with which the mechanical equivalent of heat is known does not have any significance.

As progress in statistical physics continued its course, the conviction that the second law has statistical nature gained ever more recognition. But far from all processes occurring in Nature have probabilistic character. Although the laws of the theory of probability do leave an imprint on many phenomena of Nature, it is in no way indelible, even in the frame of thermal processes. While randomization is a characteristic feature in the behavior of closed stochastic systems which are close to the state of statistical equilibrium, self-organization in Nature is not less natural than the randomization is. To sum up, the second law, unlike the first one, does not originate from any principles of a general nature.

One commonly considers both the first and second laws as not only the principles of thermodynamics but the laws of Nature as well. As for the first law, one cannot raise here any doubts whatsoever. The situation is radically different with the second law. Even if we assume both laws to apply to a large variety of quasistatic processes, there are no grounds whatsoever for extending the second law to processes involving intense material and heat flows at all. An analysis of the totality of nonequilibrium thermodynamic processes aimed at finding proofs of the validity of the second law will hardly ever become realizable.

Objections of a fundamental nature against the second law were expressed by J. Maxwell ("Maxwell daemon"). They were, however, shall we say, of an exotic nature. The second law states essentially the impossibility of realization of a thermodynamic process for conversion of heat to work without the use of a cooler.

The unflinching belief into the second law of thermodynamics seams all the more strange that its refutation was literally lying on the surface. Already in the mid-XIX century the data sufficient for its disproof were available. One of the equivalent formulations of the second law reduces actually to the statement that cooling an adiabatic system in a repetitive series of variation of its external parameters is impossible. For any single-parameter system this is indeed so. In the case of a gas, an adiabat in a (*PV*) diagram is the steeper, the smaller the specific heat capacity of the gas. Expansion of a gas in adiabatic conditions produces work, and the gas cools down. The reverse process restores the gas to its original state. But if there were a possibility to increase the specific heat capacity of the expanded gas, its compression would follow a smoother adiabat than that governing expansion. As a result of the gas recovering its volume to the original level, the work expended in compression should be less than that produced in expansion, and the original temperature would not be reached. There are no means for varying the heat capacity of a gas at constant volume, but if it is brought in thermal contact with another object (a liquid, a solid body), the behavior of this gas under variation of its volume will be governed by its total heat capacity. The adiabat in the (*PV*) diagram of a gas in contact with another object will be smoother than the one associated with the gas. Increasing the heat capacity of the contacting object will make it still smoother. As a result, expansion of a gas at the minimum heat capacity of the contacting object followed by its compression back to the initial volume at the maximum one will bring about cooling of the system.

A natural question arises here, to wit, how could one vary the heat capacity of the contacting object? Is it generally possible, and, if so, would it involve expenditure of work? And would not this expenditure of work required to change the heat capacity exceed the work done by the gas?
A body with a controllable heat capacity can be illustrated with a particular example of a long thin rod with a positive coefficient of elongation with temperature. The energy of the rod is the sum of its internal energy and the potential energy determined by the position of the center of gravity in the Earth's gravitational field. When the rod rests in horizontal position, a change in temperature resulting in the corresponding change of the length of the rod does not affect the height of its center of gravity, whereas when it is in the vertical position, the direction of the change in the height of the center of gravity of the rod depends on what point of the rod was fixed. If it is the lower end that is fixed (the rod is standing upright), an increase in rod length results in a rise of its center of gravity, while when the top end is fixed (the rod is hanging), it goes down. Accordingly, the heat capacity in the first case is larger than that in the second.

Cooling of an adiabatic system can proceed in the following way. In the initial state, the temperature of the rod is equal to that of the thermostat $T_0$. The rod is in the vertical position. The volume of the gas is minimal. In the first stage, the rod is transferred to a hanging position by turning it about the pivot. A certain amount of work is done. The temperature of the system does not change. In the second stage, the gas expands and also produces some work. The temperature of the system decreases, and the rod length contracts. In the third stage, the rod is restored to the vertical position. As a result, the center of gravity will now be lower than it was initially. The work expended in this stage is smaller than that done in the first stage. The temperature remains unchanged. In the concluding stage, the system recovers its original volume, and heat is released. Because compression of the gas occurs with the heat capacity of the system being larger than that during expansion, the work expended in compression is certainly smaller than that done in expansion. Thus, both the gas and the rod have produced work. This can occur only as a result of lowering of the system temperature. But the rod did not recover its original length. The system can return to the initial position if thermal contact with the thermostat is established.

The adiabat of the gas in thermal contact with the rod is given by the relation
$$(C + C_V)dT + PdV = 0 \tag{9}$$
where C is the heat capacity of the rod. In the second stage of the process under consideration, $C = C_{down}$, and in the fourth, $C = C_{up}$. As a result, for the temperature in the concluding stage we obtain
$$T_{fin} = T_0 \left(\frac{V_0}{V_{max}}\right)^{\frac{N(C_{up} - C_{down})}{(C_V + C_{up})(C_V + C_{down})}} \tag{10}$$
where $V_0$ is the initial volume of the gas, and $V_{max}$, its maximum volume. The exponent in Eq. (10) being positive, $T_{fin}$ is certainly less than $T_0$. Even without consulting reference books on thermophysical parameters of materials, it appears obvious that in real conditions, no matter how large is a gas expansion, the cooling effect is extremely small, but its fundamental significance cannot be overestimated. While being simple, the above example can hardly be considered as a more convincing objection to the second law than the Maxwell demon is. By the way, Internet reported recently that the Maxwell demon has been created in Japan [12].

# CONVERSION OF HEAT TO WORK

# IN A THERMALLY HOMOGENEOUS TWO-PARAMETER SYSTEM

Variation of thermodynamic characteristics of a system in the course of conversion of heat to work is a process hard to realize. In the case of solids and liquids, the dependence of thermodynamic parameters on the conditions of a process is negligibly weak. The heat capacities $C^{up}$ and $C^{down}$ differ hardly more than in the fifth or in the sixth digit. The heat capacities of gases depend much more significantly on the actual conditions of the process being conducted.

Consider a two-parameter, thermally homogeneous system of gases separated by a fixed heat-conducting partition. The independent variables of such a system are the two volumes of the gases and their common temperature. Recalling the equation of the perfect gas, the Pfaff form of such a system can be written in generally accepted notation as

$$\delta Q = CdT + \frac{N_1 T}{V_1} dV_1 + \frac{N_2 T}{V_2} dV_2, \qquad (11)$$

where C is the heat capacity of the system, which is the sum of the heat capacities of the two gases. Because the heat capacity of a gas can depend only on temperature, the form (11) is certainly integrable, and it would seem that the second law should not bring here any inconvenience. Such a conclusion would be though too hasty. Indeed, heat capacity depends not only on the properties of the substance undergoing heat exchange but on the conditions in which this heat exchange is run too. The principal idea involves the possibility of controlling the thermodynamic characteristics of a system in the course of a thermal process [13].

Consider an adiabatic system consisting of two compartments which are filled by two gases separated by a fixed heat conducting partition. The thermal contact ensures that the temperatures of both gases remain equal during a quasistatic process. To the right of the partition a horizontal cylinder with an actuating piston is located, and to the left of it, two interconnected cylinders, one of which is horizontal with an actuating piston, and the other, vertical with a heavy piston (Fig. 1a). The two actuating pistons are in fact the external parameters of the system.

With the heavy piston lying on the bottom of the vertical cylinder, the maximum possible pressure in the left-hand part of the system is determined by the weight of the heavy piston. At lower pressures, the volume of the gas in the left-hand section is unambiguously determined by the position of the left-hand actuating piston. In these conditions, motion of the left-hand actuating piston results in variation of the volume and pressure of the gas in volume $V_1$, and the Pfaff form coincides with that for two gases separated by a heat-conducting partition. The heat capacities of both gases are here equal to those at constant volume.

If the heavy piston does not rest on the bottom of the vertical cylinder, the gas pressure in the left-hand section is constant and determined by the weight of the piston. Motion of the left-hand actuating piston makes the heavy piston move too, with the volume $V_1$, temperature and pressure of the gas in it remaining unchanged. The left-hand section of the system is of potential character [14]. The heat capacity of the gas in the left-hand section is equal to that at constant pressure.

In both cases, the Pfaff form of the system has an integrating factor. But if in the course of a thermodynamic process the heavy piston is brought to the bottom and returned thereafter to the height of free motion, one can hardly assume the Pfaff form, and all the more so, its integrating factor to remain unchanged throughout the process. Neither the heat capacity of the system nor its Pfaff form are smooth functions. Nevertheless, such processes can be readily studied in a general form.

It is instructive to consider the following four-stage process. In the first stage of the process, the actuating piston in the left-hand horizontal cylinder moves to the left until the heavy piston comes to rest at the bottom of the vertical cylinder (Fig. 1b). The work done in the process is equal to the decrease of the potential energy of the heavy piston in the field of gravity, with the left-hand part of the system losing its potential character.

In the second stage, the volume of the right-hand section increases as its piston moves to the right, the process in which the work is done and the system is cooled as a whole (Fig. 1c). The gas pressure in the left-hand section decreases. The cooling is continued to the extent where the gas pressure in the left-hand section with the left-hand actuating piston reaching the original position has not yet recovered its original level. If the cooling was deep enough, the energy of the gas in the left-hand section transfers practically completely to the right-hand section, and, subsequently, is converted to work. In the limiting case of unrestricted expansion of the gas in the right-hand section, the internal energy of the gas in both sections of the systems is converted completely to work.

After this, the actuating piston in the left-hand section is returned to its original position (Fig. 1d). The work expended in returning the actuating piston is notoriously smaller than that done in the first stage. In these conditions, the heat transferred to the system in the process may be neglected. In the concluding stage, the right-hand actuating piston resumes its original position, and the system as a whole heats (Fig. 1e). In the beginning of the stage of compression, the structure of the system differs from that in the stage of expansion in the volume of the gas confined in the left-hand section. The gas heat capacity being independent of volume, the thermal characteristics of the system under compression are identical with those under expansion. Heating of the gas in the left-hand section takes place at constant volume, and the right-hand section passes through the same sequence of states as in the second stage, but in reverse order. After the gas in the left-hand section has reached the original pressure due to the heating, this part of the system becomes potential again, with the corresponding radical change of its thermal parameters. Further heating of the system occurs with the gas in the left-hand section maintained at constant pressure rather than at constant volume. The growth of pressure and temperature, as well as of the work expended in compression, slows down as the volume of the right-hand part decreases. After the left-hand piston (in the third stage) and the right-hand one (in the concluding stage) have resumed their original positions, the temperature has not reached its initial level.

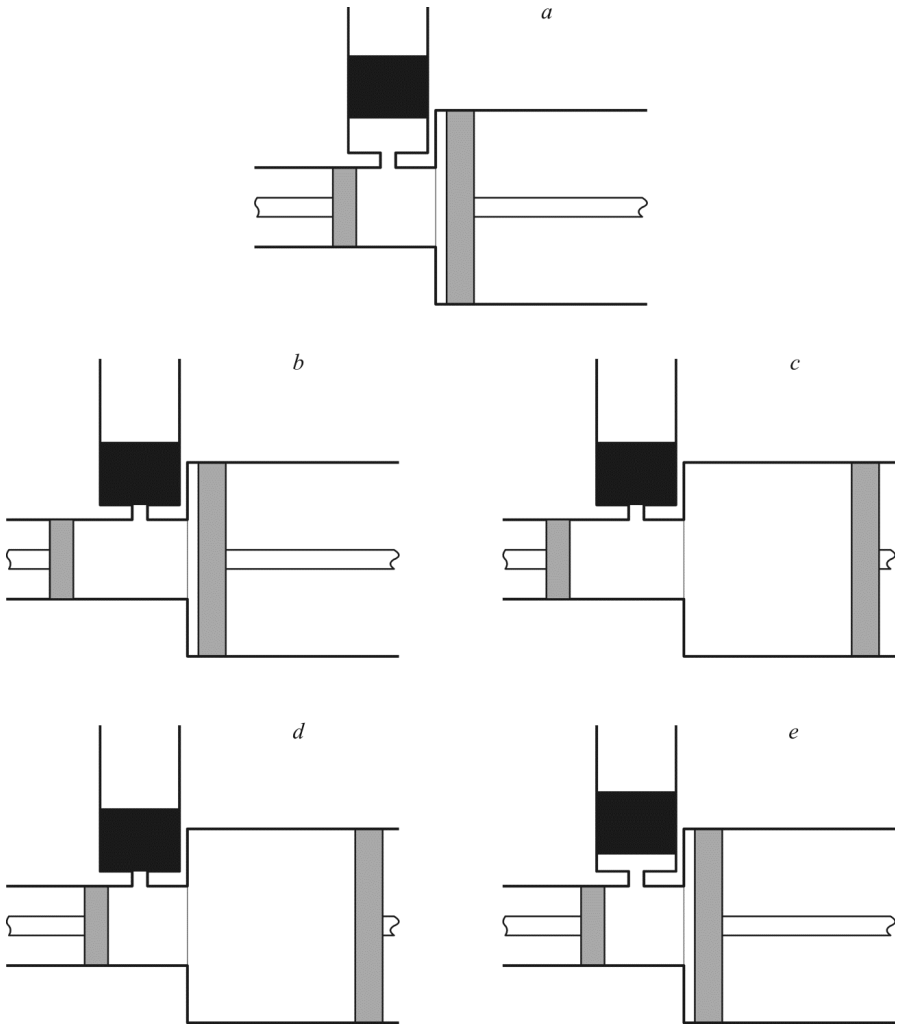

Fig. 1. State of the system in different stages of the process.

(a) Original position; (b) end of the first stage; (c) end of the second stage; (d) end of the third stage;

(e) end of the concluding stage.

To sum up, the return of the external parameters of the system to their original values is accompanied by its cooling and production of work, the end result that comes into conflict with one of many equivalent formulations of the second law of thermodynamics. To restore all thermodynamic characteristics of the system to their original values, one has now only to establish thermal contact with the thermostat which maintains the initial temperature.

As a particular example, consider the case of a perfect monatomic gas. The numbers of molecules of the gases $N$ in both parts of the system and their initial pressures $P_o$ and volumes $V_o$ will be assumed equal. The changes of the parameters of the system induced by a change of only one external parameter are specified by the well-known relations describing an adiabatic process [5]

$$T^{\gamma}P^{1-\gamma}=\text{const}, \qquad TV^{\gamma-1}=\text{const}, \qquad PV^{\gamma}=\text{const}. \qquad (12)$$

Heat exchange between the two sections of the system culminates in an increase of the heat capacity of each section. If the volume of the contacting section remains constant, the heat capacity of the varying section doubles, while in the case of constant pressure the heat capacity of the varying section increases through addition of that of the contacting section at constant pressure. Accordingly, the exponent of the "adiabat" in the first case will be now $\gamma_1 = 1 + 1/2c_v$, and that for the second, $\gamma_2 = 1 + 1/(2c_v + 1)$, where $c_V$ is the heat capacity of the gas at constant volume per one molecule. For a monatomic gas, $c_V = 3/2$, which leads to $\gamma_1 = 4/3$, and $\gamma_2 = 5/4$.

Consider first the case with the exponent of the "adiabat" remaining constant in each stage of the cyclic process. In the first stage, the left-hand actuating piston moves to the left until the heavy piston comes to rest at the bottom of the vertical cylinder. The thermal characteristics of the system do not change.

The increase of the volume of the gas in the right-hand section of the system in the second stage and its cooling can be found from the condition of the gas pressure in the left-hand section recovering its initial value at the end of the third stage (after the left-hand actuating piston has regained its initial position). The exponent of the "adiabat" undergoes a change at the transition from the third to the concluding stage. Obviously enough, the temperature of the system at the end of the third stage, $T_3$, is $V_0/V_3$ times lower than its initial value; $V_3$ is the gas volume at end of the third stage. The temperature of the system in the beginning of the third (or at the end of the second) stage, $T_2$, can be found from the condition $T_2V_0^{1/3} = T_3V_3^{1/3}$. This gives $T_2 = (V_3/V_0)^{4/3}T_0$. The volume of the right-hand section at the end of the second stage $V_2$ is specified by a similar condition. We finally obtain $V_2 = (V_0/V_3)^4 V_0$. The final temperature $T_f$ can be derived from the condition $T_fV_0^{1/4} = T_3V_2^{1/4}$. In the end we come to $T_f = T_0$. The extent of the cooling of the system is specified by the magnitude of $V_2$. For $V_2 < V_0(V_0/V_3)^4$, the transition from the constant-volume to constant-pressure gas condition in the left-hand section occurs already in the third stage of the process. For $V_2 > V_0(V_0/V_3)^4$, this transition takes place in the concluding stage. If $V_2 < V_0(V_0/V_3)^4$, the system having been cooled insufficiently in the second stage, compression in the left-hand section raises the gas pressure to its initial level, with the system acquiring again potential characteristics. Further motion of the left-hand actuating piston results only in sliding up of the heavy piston while not affecting in any way the thermal parameters of the system. The temperature $T_c$ at the transition to constant gas pressure in the left-hand section is found from the condition $T_cP_0^{-1/4} = T_3P_2^{-1/4}$, where $T_2$ is the temperature of the system at the end of the second stage, and $P_2$ is the gas pressure in the left-hand section at the end of the second stage: $P_2 = NT_2/V_0$. Because $P_0 = NT_0/V_0$, we have $T_c = T_2(T_0/T_2)^{1/4}$.

At the transition to the concluding stage, the temperature of the system $T_4 = T_c$. The temperature in the end is derived from the condition $T_fV_0^{1/4} = T_4V_2^{1/4}$. We finally come to $T_f = T_0$. This actually means that if the exponent of the "adiabat" is constant, no cooling occurs in the concluding stage. For $V_2 > V_0(V_0/V_3)^4$, the situation becomes radically different. In this case, the return of the actuating piston to the original position and the transition to the concluding stage take place at the gas pressure in the left-hand section lower than its original level. The temperature $T_c$ of the transition to constant gas pressure in the left-hand section can be found from the equation of state of an ideal gas: $T_c = P_0V_3/N = T_0V_3/V_0$, and the volume of the gas in the right-hand section of the system, $V_c$, from the condition $T_cV_c^{1/3} = T_3V_2^{1/3}$. The temperature the system reaches in the end is given by the condition $T_fV_0^{1/4} = T_cV_c^{1/4}$. Thus, $T_f = (V_3/V_0)^{3/4}T_0$.

Significantly, the final temperature is not an analytic function of the system parameters. If cooling in the second stage of the process was not intense enough, the final temperature recovers its original value, irrespective of the extent of cooling. If it was deep enough, the final temperature decreases, but not under unlimited gas expansion in the second stage, where it is fairly obvious, but immediately after the condition $V_2 > V_0(V_0/V_3)^4$ has been met. Further cooling of the system in the second stage does not affect the final temperature. Such a "bistability" of the result of the process, rather than of a state, has not been earlier reported in thermodynamics.

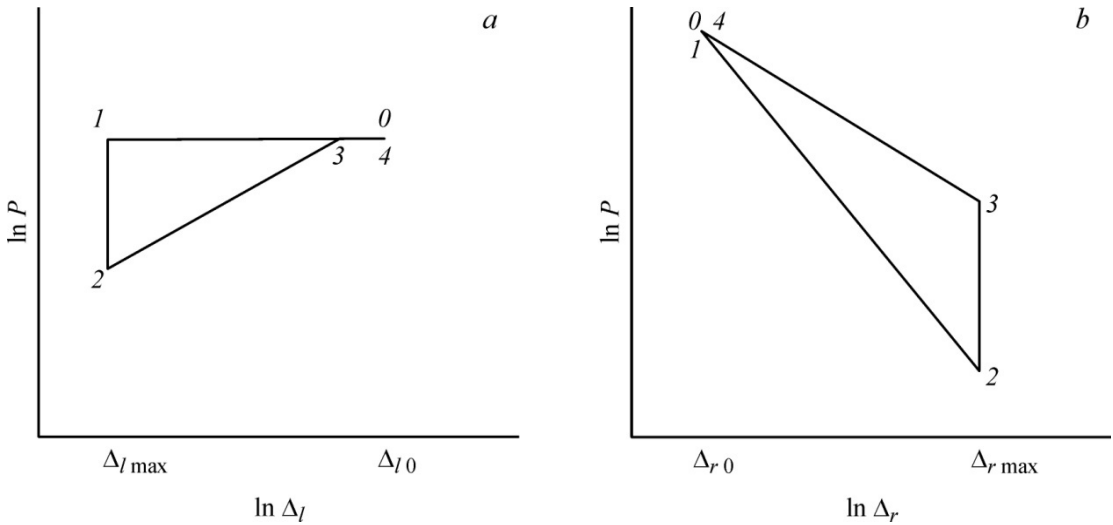

Fig. 2. Pressure--actuating piston displacement diagram plotted for $V_2 < V_0(V_0/V_3)^4$.
(a) Left-hand piston; (b) right-hand piston.

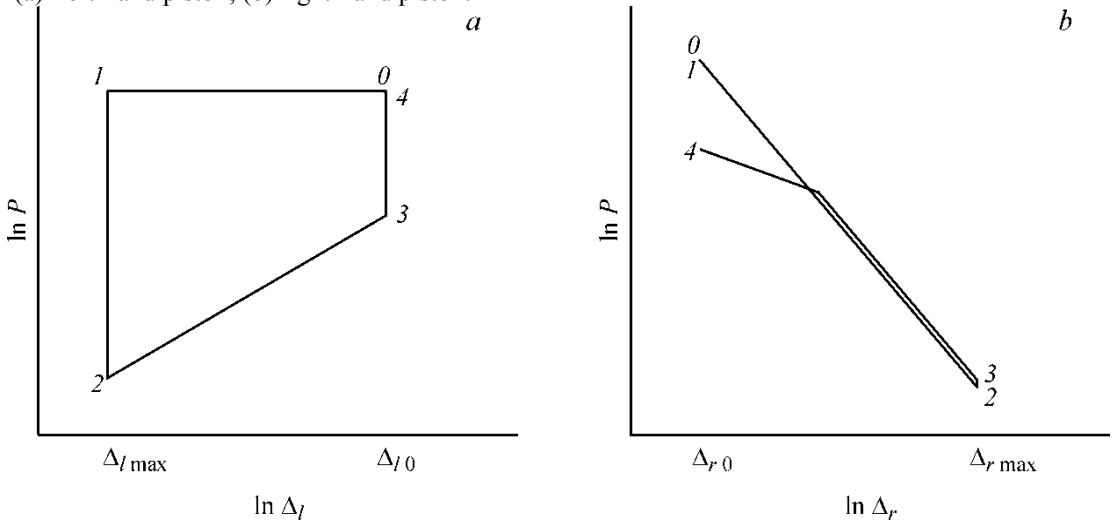

Fig. 3. Pressure--actuating piston displacement diagram plotted for $V_2 > V_0(V_0/V_3)^4$.

(a) Left-hand piston; (b) right-hand piston.

Figure 1 visualizes the sequence of states the system assumes in the course of the process for $V_2 > V_0(V_0/V_3)^4$, and Figs. 2 and 3 illustrate the "pressure--actuating piston displacement" diagrams drawn for different values of the system parameters. The position of the right-hand piston defines unambiguously the volume of the gas in the right-hand section of the system. For the left-hand section, such a mutual one-to-one correspondence sets in only at pressures below the original value. Thermodynamic curves (straight lines if plotted in the log scale) specify the extent of the work done. The work done by the left-hand piston in the cycle is always positive. For $V_2 < V_0(V_0/V_3)^4$, the work done by the right-hand piston during the cycle is negative and equal in absolute magnitude to that of the left-hand one. In the limiting case of $V_2 \to \infty$, the work done by the right-hand piston in the cycle is positive. For $V_2 < V_0(V_0/V_3)^4$, the cycle is closed for all the variables, and for $V_2 > V_0(V_0/V_3)^4$, only for the external parameters.

# CONCLUSION

The above examples of total conversion of heat to work were based on the energy conservation law and, by default, on the unquestionable statement that "in bodies brought in thermal contact, heat transfers by itself from the more heated body to the less heated one", which is in no way equivalent to the second law. The recovery of the original state of the system by heating is irreversible. Therefore, strictly speaking, all the above processes do not fit within the bounds of thermostatics. The way out is, however, so simple and straightforward that its validity can hardly raise any doubts.

Entropy is a measure of the extent of order in a stochastic system obeying the laws of the theory of probability. The theory of probability can be extended only to an extremely limited range of stochastic systems and processes. Indeed, even in the simplest cases the probability is defined as the limit to which tends the ratio of the number of realizations of an event to the total number of attempts as this number increases indefinitely. In a general case, such a limit may not exist at all. In a general case, statistical systems or processes feature a random behavior. There is no reliable basis for their study. In the Mathematical Encyclopedia [4] there is even no definition of the concept of "chaos". The probability theory can be used only if the states of a system can be specified by their probabilities. In single-parameter systems, this condition is met always. A change of an external parameter or of temperature brings about a change not only of the probabilities but of the sampling space itself. In this sense, probabilities are a posteriori (conditional) rather than a priori quantities. It is entropy that serves as a measure of the extent to which a system is ordered. In equilibrium, thermodynamic entropy and entropy in the theory of probability essentially coincide. In many-parameter systems, a change of any external parameter entails a change of the characteristics not only of the corresponding part of the system but of all its parts. Many-parameter systems in the state of equilibrium can be thermally inhomogeneous as well. Such systems are, generally speaking, nonintegrable, and, therefore, do not have entropy at all. But in the case of the heat capacity pressure dependent, thermally homogeneous systems are nonintegrable too and, as we have seen, allow total conversion of heat to work.

Even staying within the limits of quasistatic processes, macroscopic bodies may not have entropy either by their nature, for example, if they are thermally inhomogeneous, or in respect to the process in which the thermodynamic parameters of the system change. It is the second law and the principle of maximum entropy that provide the stability of local thermodynamic equilibrium and equal temperatures of all parts of the body at equilibrium. Irrespective of whether a body divided in parts is in equilibrium or not, each of its parts is a single-parameter system, and in respect to it all of the argumentation based on the second law remains applicable. Under local equilibrium, each part of the system has both its specific temperature and its entropy. In these cases, the second law is itself a consequence of the corresponding Pfaff form being integrable. The system as a whole, however, has neither a specific temperature nor entropy. Although the thermodynamic entropy and the entropy dealt with in the theory of probability have different definitions, thermodynamic entropy has a meaning only when applied to equilibrium systems, where the distribution function and the distribution of probabilities actually coincide.

If there are no adiabatic partitions, equalization of temperatures among different parts of a body can be explained in terms of the statement presented above in quotation marks, or of its equivalent. The first law alone is insufficient both for determination of temperature, because of the law having unambiguous character, and for designing a method for its measurement. How should the "second law" read remains an open question, but certainly it cannot have the form assumed by A. Einstein.

In thermodynamics of single-parameter systems, the only way to reduce the entropy of a system is by cooling it. A remarkable property of both the temperature and the entropy is that in an adiabatic system subjected to any cyclic variation of external parameters, irrespective of whether they are reversible or not, they can, as required by the universally accepted concepts, only increase. But the temperature of a many-parameter system can decrease not only under cooling but as a result of a cyclic variation of the external parameters as well. A decrease of temperature provides perhaps a more universal proof of "ordering" in the system as a whole than that of entropy, because in a general case this entropy does not exist. But even in the situations where thermodynamic entropy does exist, it is determined to within an arbitrary constant. The "third law", in contrast to the zero one, cannot be ranked among the fundamental principles.

Cooling of an adiabatic thermodynamic system resulting from a cyclic variation of its external parameters may be considered as a process of self-organization. While there are no criteria of a general nature characterizing the extent of chaos in a system, cooling of all its parts may certainly be considered as strong evidence for self-organization it has undergone [15--17].
Self-organization is one of unquestionably existing while certainly mysterious manifestations of Nature. Life would be inconceivable without self-organization. But inanimate nature also reveals signs of self-organization. A graphic demonstration staggering our imagination is provided by tornadoes (if tearing trees with roots out of the ground, destruction of houses and loss of people may be regarded as a result of self-organization). It is a common belief that self-organization is possible in open systems only (see [15] and references therein). A closed many-parameter system is an array of statistical systems, with each constituent part possessing its own sampling space. Individual parts of a system may be treated as open with regard to its other parts. Even in closed two-parameter systems, a quasistatic process is accompanied under certain conditions by self-organization, with the external parameters providing control.

Using the possibility of controlling thermodynamic characteristics with the aim of increasing the organization of a thermodynamic system does not certainly exhaust the ways on which such an increase can be reached. One can employ for this purpose external fields, as well as inertial forces evolving in the course of macroscopic motion of media. As a result, even at equilibrium, in a concomitant coordinate system there may appear fairly large gradients of gas pressure and temperature. Temperature uniformity is a characteristic feature only in the case of absence of force fields.

It is customarily assumed that the equilibrium distribution function of a gas in the gravity force field is defined by the barometric relation. The temperature of a gas is independent of altitude. It is well known, however, that not only the pressure of the air in the Earth's atmosphere but its temperature as well fall off with altitude. This is usually assigned to the atmosphere being not at equilibrium. Gas convection in the gravity force field may produce, because of atmospheric pressure being dependent on altitude and the low thermal conductivity of the gas, temperature gradients. This process, combined with heating of the air by the Earth's surface, should result in a decrease of the temperature with altitude, which will be limited by the condition of stability of mechanical equilibrium (adiabatic stratification of the atmosphere). The temperature gradient may be as high as about ten degrees per kilometer [1].

Under thermodynamic equilibrium and with no gravitational field present, the internal energy, pressure and temperature of the gas should not depend on coordinates. All elements of the gas are likewise at equilibrium. Each element is characterized by its own specific thermodynamic parameters. (Local thermodynamic equilibrium.) Not only the pressure and temperature, but the specific entropy as well are coordinate-independent. The gas having a low thermal conductivity, the motion of small elements of mass may be assumed to be adiabatic. At equilibrium, mixing does not affect the state of the system, which is possible only if specific entropy is coordinate independent. The gas is assumed to contain molecules of one species only. Both the parameters of the system as a whole and those of the elements of mass satisfy the energy conservation law, which is specified by the well known Pfaff form.

Being the integrating divisor of the Pfaff form. temperature is a function of its internal energy and volume. But since temperature, as the integrating divisor of the Pfaff form, depends on internal energy, at thermodynamic equilibrium the temperature cannot be independent of altitude if the internal energy depends on altitude in the gravity force field.

In a general case, internal energy is not related in any way with temperature. Unlike temperature, internal energy is not a measurable quantity and is specified to within an arbitrary constant. Variation of the altitude of an element of mass in a gravitational field at a fixed volume changes significantly the internal energy of the gas while not affecting in any way its temperature. Both the internal energy and volume are independent variables which fully characterize the state of an equilibrium thermodynamic system. The internal energy and temperature of an element of gas mass may become closely related only if both variables defining its state vary.

It would seem that if the dependence of internal energy on altitude is known, it should not be difficult to derive that of the temperature. To do this, however, one should know the value of the heat capacity of the gas, but its heat capacity depends on the actual conditions of heating or cooling. Indeed, in the presence of the gravity force field not only internal energy but pressure too depend on altitude. Self-diffusion in a gas is realized through transposition of various elements of mass, a process which does not change the gas state at equilibrium. Transposition of elements of mass in altitude involves a change of pressure in them in this case, and, hence, of their volumes and temperatures.

Determination of the equilibrium distribution function from the condition of maximum entropy is a typical isoperimetric problem. The assumed constancy of undefined Lagrange factors [5] accounts usually for the possibility of exchange of energy among different elements of mass while not allowing for the possibility of their displacement, and, by default, brings about independence of temperature on altitude. This corresponds to maximum entropy, provided not only the internal energy and the number of particles in the system are conserved but the free motion of molecules is forbidden. This exclusion can be removed only by allowing independence of specific entropy on coordinates. In this case, all of the system's volume is accessible to each element of mass and each molecule of the gas, with transposition of different elements of mass affecting now neither the energy nor the entropy of the system.

Written in common notation, the equilibrium temperature gradient is given by the following relation

$$\frac{dT}{dz} = \frac{\gamma-1}{\gamma} \frac{T}{P} \frac{dP}{dz} = -\frac{\gamma-1}{\gamma} \frac{mg}{k} \qquad (13)$$

It coincides with the boundary of convective stability [1]. A faster drop of the temperature with altitude would result in mechanical instability of the system and development of convection, and a slower one, to decrease of the entropy. Not only the mechanical but thermodynamic equilibrium as well are realizable at the maximum possible falloff of the temperature with altitude allowed by the condition of mechanical equilibrium, and at maximum entropy. The existence of gradients of pressure and temperature in the equilibrium state does not translate into initiation of material or heat flows. Only in the state of neutral mechanical equilibrium and maximum entropy, will mixing of gas masses affect neither the energy nor the entropy of the system.

A linear decrease of temperature with altitude does not change significantly either the Euler equation or the form of the equilibrium probability distribution function. The only thing one has to do is to replace $T$ with $T_0 - (z-z_0)(\gamma-1)mg/\gamma k$, where $T_0$ is the temperature at level $z_0$. The dependence of temperature on altitude accounts for high enough $z$ for the sharp gas boundary, which corresponds to the temperature becoming zero. However, while the temperature falls off linearly with altitude as one approaches the boundary, pressure decreases exponentially with an indefinitely increasing exponent.

One will hardly succeed in detecting such extremely small temperature gradient of the air stemming from the force of gravity in laboratory conditions. The force of gravity is too weak for that. Although the main concepts underlying thermodynamics were defined for stationary bodies and have a meaning in the concomitant coordinate frame, thermodynamic equilibrium, at any rate on a local scale, can obtain in the case of moving bodies too, in particular, in centrifuges. Using up-to-date centrifuges and heavy gases, one would certainly be able to detect nonuniformity of temperature in equilibrium state.

Vortex tubes offer a unique possibility for the study of the effect of acceleration on the equilibrium temperature of a gas. For a tube of about one cm in radius and injection of gas with a transonic velocity, the accelerations which can be typically reached in a vortex flow exceed by six orders of magnitude the free-fall gravitational acceleration. The temperature gradients evolving in such a huge acceleration may be as high as ten degrees, and not per one kilometer but per one millimeter. Accordingly, the difference of temperature between the "hot" and "cold" gas flows leaving the tube should rise to many tens of degrees [18].

The possibility itself of the gravity being capable of affecting the gas temperature in thermodynamic equilibrium is in direct conflict with the generally accepted formulations of the Second Law of Thermodynamics [1]. This statement is indisputable. The linear dependence of temperature on altitude is one more irrefutable proof of the need of revising the content of the formulations of the Second Law.

Meteorology was one of the subjects in the realm of scientific interests of Ya. I. Frenkel. In discussing an aspect of meteorology, Yakov Il'ich loved to repeat: "A cloud is not just an object. Cloud is a process". Actually, any macroscopic object is linked inseparably with a process occurring in it. But in a cloud this manifests itself most clearly.

The Fathers of thermodynamics and its adepts, without even making a try at getting to the essence of things, had not been late in expressing their categorical judgment about the processes that can occur with these objects. Robed into a pseudo-scientific form [6, 19], the judgment was announced to be a Law of nature. But this was essentially nothing more than a reverence in support of the decision of the French Academy of Sciences. As a result, the engineering genius and the public opinion had been disoriented for many years.

In quasistatic processes, there are no limitations on the possibility of total conversion of heat into work. But they are of no interest in the area of applications, because their power is zero. Realistic systems designed for conversion of heat to work are linked inseparably to the motion of the working substance, therefore investigation of their properties lies beyond the framework of the present-day thermodynamics.

# ТЕРМОДИНАМИКА
## ОСНОВНЫЕ ПРИНЦИПЫ

*Г.В.Скорняков*

Физико-технический институт им. А.Ф.Иоффе РАН, Санкт-Петербург, 194021, Россия;
e-mail: skorn@ioffe.ru

ПРЕДИСЛОВИЕ

Этот текст – не для ознакомления с теорией тепловых процессов. Основные принципы теории имеет смысл обсуждать лишь с компетентными людьми. Автор надеется на знание читателем этой теории в объёме программы общего курса физики физического факультета университета [1]. Стимулом для написания книги явилась установленная автором несостоятельность второго начала термодинамики как закона природы [2].

Тепловые процессы не только являются предметом научных исследований, но и пронизывают буквально всё наше существование. Термодинамику принято считать научной базой теплоэнергетики, а преобразование тепла в работу – основной проблемой термодинамики. Преобразование тепла в работу с помощью различного рода тепловых машин получило широкое промышленное применение задолго до возникновения термодинамики. Основной целью термодинамики с самого начала был поиск наиболее эффективных методов этого преобразования. Но именно термодинамика явилась непреодолимой монолитной стеной между поистине неисчерпаемым количеством окружающей нас тепловой энергии и возможностью её практического использования.

Отказ от второго начала, как от универсального закона природы, казалось бы, должен привести к пересмотру всех сложившихся представлений об окружающем нас мире. В действительности положение не столь трагично. Дело в том, что подавляющее большинство (если не все) рассмотренных в термодинамике процессов преобразования тепла в работу относится к однопараметрическим системам. Второе начало для них безусловно справедливо. Но если число внешних параметров системы превышает единицу, применимость к ней второго начала требует доказательства. Так что второе начало в его общепринятом понимании – не более чем распространение банального на область неизвестного.

Второе начало – глубочайшее, но не единственное заблуждение в установившихся представлениях о характере тепловых процессов. Другим является фактическое отождествление функции распределения неравновесной системы с распределением вероятностей. И если начала термодинамики в формулировке А.Эйнштейна содержат лишь «техническую» неточность, то невозможность отождествления функции распределения неравновесной системы с распределением вероятностей носит концептуальный характер и не только лишает смысла любые попытки распространения сложившихся в термодинамике понятий на процессы в неравновесных системах, но делает проблематичной саму возможность их исследования в рамках современной математики.

Принято считать, что "физическую сущность" основных понятий термодинамики раскрывает статистическая физика. Молекулярно-кинетическая теория призвана не только разъяснить понятия феноменологической теории, какой с самого начала представляла собой термодинамика, но и определить границы их применимости.

Термодинамика не формулирует уравнений, определяющих эволюцию состояния термодинамической системы во времени. Она устанавливает лишь общие принципы, в рамках которых эта эволюция может происходить. Такой характер термодинамики создавал чувство неудовлетворенности у многих исследователей. Ряд установленных в термодинамике понятий (температура, энтропия) проник в описание заведомо неравновесных процессов. Более того, даже формулировка основных уравнений гидродинамики оказалась невозможной без их использования. Многие пришедшие из термодинамики понятия настолько въелись в сознание, что их применение зачастую происходит по умолчанию.

Анализу основных понятий термодинамики, лежащих в её фундаменте категорий, и посвящён настоящий текст. Особое внимание уделено математическим основам используемых в термодинамике представлений. Никаких конкретных задач в нём не рассматривается.

## ВВЕДЕНИЕ

Начало XX века ознаменовано двумя событиями, в корне изменившими характер научных представлений о материальном мире, природе пространства и времени – возникновением квантовой теории и созданием теории относительности. Теория относительности связала воедино пространство и время, устранила известное противоречие между классической механикой и электродинамикой, а затем, после создания общей теории относительности, привела к современным представлениям о связи вещества, пространства и времени, лежащим в основе наших знаний о структуре и эволюции Вселенной. Ещё более радикальные изменения в характер научных представлений внесла квантовая теория, установившая качественно иную природу закономерностей микромира по сравнению с тем, что стало привычным за многие годы исследования макромира.

Однако общей характерной чертой этих двух величайших революций в истории естествознания было то, что они не внесли заметных изменений в установленные ранее законы, а лишь ограничили область их применимости. Более того, законы классической физики оказались предельным случаем новых законов, справедливым при малых по сравнению со скоростью света изменениях скоростей и при больших по сравнению с постоянной Планка изменениях действия. Такое соотношение старой и новой теории, в духе принципа соответствия, является необходимым признаком объективности научного знания и его непреходящего значения.

Один из основных разделов классической физики – термодинамика. Возникновение термодинамики как науки относится к началу XIX века. Не будет преувеличением сказать, что основы термодинамики заложены Сади Карно, впервые сформулировавшим фундаментальное положение, получившее впоследствии наименование второго начала термодинамики. Первое начало термодинамики было установлено значительно позже, в середине XIX века. Термодинамика возникла и продолжает существовать как феноменологическая теория, опирающаяся непосредственно на экспериментальные данные, теоретическим обобщением которых она является. Молекулярно-кинетическая теория подвела теоретическую базу под основные феноменологические положения термодинамики. С созданием квантовой теории построение стройного здания термодинамики, опирающегося на прочный фундамент данных опыта, было завершено.

Термодинамику часто сравнивают с геометрией. Подобно геометрии она построена на аксиоматической основе По широте охвата явлений термодинамика не имеет себе равных среди физических теорий. В этой связи А.Эйнштейн писал [3]:

"Теория производит тем большее впечатление, чем проще её предпосылки, чем разнообразнее предметы, которые она связывает, и чем шире область её применения. Отсюда глубокое впечатление, которое произвела на меня классическая термодинамика. Это единственная физическая теория общего содержания, относительно которой я убежден, что в рамках применимости её основных понятий она никогда не будет опровергнута (к особому сведению принципиальных скептиков)".В чем же усматривал А.Эйнштейн наиболее привлекательные черты термодинамики? Вспоминая о создании теории относительности, он подчёркивал [3]:

"Только открытие общего формального принципа может привести нас к надежным результатам. Образцом представлялась мне термодинамика. Там общий принцип был дан в предложении: законы природы таковы, что построить вечный двигатель (первого и второго рода) невозможно... Общий принцип специальной теории относительности содержится в постулате: законы физики инвариантны относительно преобразований Лоренца... Это есть ограничительный принцип для законов природы, который можно сравнить с лежащим в основе термодинамики ограничительным принципом несуществования вечного двигателя."

Именно термодинамика, точнее – термодинамика излучения, явилась первым мостом между классической физикой и квантовой. Только непоколебимая уверенность в безусловной справедливости термодинамики могла стать фундаментом для коренного изменения основных физических представлений, которого с самого начала потребовала квантовая теория. Триумф квантовой теории лишь укрепил эту уверенность.

И в то же время неоднократно отмечалось, что последовательное проведение точки зрения, основанной на законах термодинамики, приводит к разительным противоречиям с наблюдаемой картиной мира, и даже с самим фактом нашего существования. Естественно возникает вопрос, не являются ли эти противоречия результатом использования законов термодинамики за пределами их применимости?

В ряду физических дисциплин термодинамика занимает особое место. Характер термодинамического описания коренным образом отличается от характера других физических теорий – механики, электродинамики (в той мере, в какой в них можно пренебречь тепловыми процессами). В основе физической теории обычно лежит понятие состояния изучаемого объекта. Предметом теории является определение состояния и изучение процесса его изменения во времени. Совершенно иначе ставится вопрос в термодинамике.

Термодинамика изучает макроскопические тела, состоящие из огромного числа атомов и молекул, но описывает их состояние крайне ограниченным числом независимых переменных, доступных экспериментальному определению. В случае газов таких переменных всего две. Газы – удобный пример наглядного применения основных принципов термодинамики, тесно связанного со статистикой. Само понятие объекта в термодинамике неотделимо от условий, в которых он существует.

В термодинамике оказываются тесно переплетёнными физические величины различной природы. Важную роль играют внешние параметры, изменение которых непосредственно связано с совершаемой системой (или над ней) работой [1]. В случае газов в замкнутом сосуде внешним параметром служит объём. Другой важной характеристикой являются внутренние параметры, определяющие взаимодействие системы с внешней средой. В случае газов внутренним параметром служит давление. И объём, и давление газа определяются внешними условиями. Более того, само отнесение параметров системы к внешним или внутренним этими условиями и определяется. Можно исследовать процессы в газе как при постоянном объёме, так и при постоянном давлении. В последнем случае объём становится внутренним параметром, а давление – внешним. Так что лучше говорить о термодинамических переменных системы, не уточняя, какие из них внешние, а какие – внутренние [1].

Определяющие термодинамическое состояние системы переменные суть усреднённые за более или менее значительный промежуток времени характеристики сложной статистической системы и потому заданы не точно, а подвержены флуктуациям. Вблизи критических точек уровень флуктуаций достигает макроскопических величин, и основные предпосылки термодинамики, за исключением закона сохранения энергии, неприменимы.

Далеко не все макроскопические процессы являются термодинамическими. Процессы в статистически неравновесных системах заведомо к ним не относятся. Но одним из основных в термодинамике является понятие теплового равновесия, согласно которому изолированная система при фиксированных внешних условиях независимо от начального состояния в результате релаксации в конечном итоге приходит в состояние, которое в дальнейшем не меняется. Это утверждение иногда называют нулевым началом термодинамики.

По сути дела предметом термодинамики является изучение поведения макроскопических тел в предельном случае медленного изменения определяющих их состояние переменных. Такие квазистатические процессы обратимы [1]. Так что то, что принято называть термодинамикой, в действительности представляет собой термостатику. В рамках термостатики отсутствуют тепловые и материальные потоки.

## ТЕРМОДИНАМИКА И СТАТИСТИКА

Введение параметров, определяющих состояние термодинамической системы, лишь чисто внешне роднит термодинамику с другими физическими теориями. В результате задания параметров системы фиксируется не одно, а множество состояний, соответствующих заданному набору параметров. Термодинамические параметры определяют не одно конкретное состояние из этого множества, а лишь выборочное пространство [4], которому они принадлежат. В случае газов состояние описывается точкой в фазовом пространстве системы. В термодинамическом равновесии неизменными остаются лишь макроскопические параметры. Что же касается фазовой траектории в пространстве микроскопических переменных, то её конкретный вид никак не сказывается на макроскопических параметрах.

Движение молекул газа подчиняется законам классической механики. Они образуют динамическую систему с интегральным инвариантом (теорема Лиувилля [1]). Крайне запутанный характер движения молекул наталкивает на применение для его описания теории вероятностей. Это тем более оправдано, что классическая механика лишь приближённо описывает движение молекул. Последовательный квантово-механический подход [5] не только согласуется с классическим, но и позволяет устранить в нём внутренние противоречия (парадокс Гиббса). Следует, впрочем, иметь в виду, что теория вероятностей – отнюдь не всеобъемлющая теория стохастических систем и процессов. Не всякая случайная величина имеет каким-либо образом определённую вероятность.

Установление полного статистического равновесия при заданных параметрах системы требует значительного времени. Для этого все молекулы газа должны успеть эффективно провзаимодействовать между собой. Если бы термодинамические категории были применимы только к таким системам (а строго говоря это именно так) термодинамика не имела бы того значения, какое она приобрела почти за два века своего существования. Но практически молекулы газа эффективно взаимодействуют лишь с молекулами, находящимися от них на расстоянии порядка нескольких длин свободного пробега. В макроскопическом объёме с линейными размерами порядка нескольких длин свободного пробега достаточно полное равновесие устанавливается за времена порядка нескольких времён свободного пробега. Такие системы принято называть локально равновесными. Именно существование локального равновесия обусловливает возможность распространения понятий термодинамики на более или менее неравновесные системы, и в то же время кладёт предел абсолютизации начал термодинамики.

## ОДНОПАРАМЕТРИЧЕСКИЕ СИСТЕМЫ

Свойства системы определяются не только веществом, из которого она состоит, но и тем, каким образом она отделена от окружающей среды. Характерные черты "перегородки" могут быть различными. Если система помещена в неподвижный теплонепроницаемый сосуд, состояние внешней среды никакого влияния на процессы в ней не оказывает. Если стенки сосуда теплопроводны, в условиях равновесия температура системы равна температуре внешней среды. А если элементы "перегородки" могут свободно перемещаться, давление в системе совпадает с давлением во внешней среде. Последнее особенно существенно для газов. Идеальный газ даёт возможность наглядно проиллюстрировать все основные термодинамические категории.

Одной из основных термодинамических категорий является количество тепла $Q$. Первое начало термодинамики известным образом связывает малое изменение количества тепла в термодинамической системе с изменением её внутренней энергии и произведённой над ней (или ею) работой. Если в качестве внешнего параметра избрать объём, оно имеет вид:
$$\delta Q = dE + PdV, \qquad (1)$$
где $\delta Q$ - переданное газу количество тепла, $P$ - давление газа, $dV$ - изменение его объёма. Но изменение количества тепла не оказывается при этом дифференциалом какой-либо функции. Это и обусловливает саму возможность преобразования тепла в работу в циклическом процессе, используя нагреватель и холодильник. В случае описания системы лишь одним внешним параметром малое изменение количества тепла, хотя и не является полным дифференциалом, имеет интегрирующий множитель (делитель). Неотрицательную функцию внутренней энергии и внешнего параметра, являющуюся интегрирующим делителем пфаффовой формы (1) и принято называть абсолютной температурой.

Энергия и температура газа тесно связаны:
$$dE = C_V dT, \qquad (2)$$
где $C_V$ - теплоёмкость при постоянном объёме. $C_V = N c_v$, здесь $N$ - число молекул, $c_v$ - теплоёмкость в расчёте на одну молекулу. Так что закон сохранения энергии может быть записан в виде:
$$\delta Q = C_V dT + PdV. \qquad (3)$$
Здесь независимыми переменными служат температура и объём. В качестве независимых переменных можно выбрать температуру и давление. С учётом уравнения состояния идеального газа: $PV=NT$ (постоянная Больцмана равна единице), первое начало имеет вид:
$$\delta Q = C_P dT - VdP, \qquad (4)$$
где $C_P$ - теплоёмкость при постоянном давлении. $C_P = N c_p$, $c_p = c_v + 1$. Теплоёмкость при постоянном давлении существенно больше теплоёмкости при постоянном объёме, поскольку при нагреве в соответствии с уравнением состояния газа его объём увеличивается, на что должна быть затрачена энергия.

Первое начало может быть записано и в виде:
$$\delta Q = TdS, \qquad (5)$$

где $S$ - термодинамическая энтропия, - функция, дифференциал которой равен правой части уравнений (1) или (3), делённой на температуру. Следует иметь в виду, что общего метода отыскания интегрирующего множителя нет, а в тех случаях, когда интегрирующий множитель существует, он определён с точностью до множителя, равного произвольной функции энтропии. Соответственно, и энтропия не может быть однозначно определена. Важен лишь сам факт их существования. Несмотря на неоднозначность своего определения энтропия равновесной системы при любом способе её определения может служить одной из независимых переменных наряду с объёмом, давлением и температурой. При любом определении температуры энтропия может быть определена лишь с точностью до произвольной постоянной. Что же касается температуры, для её однозначного определения достаточно иметь хоть один прибор, показания которого зависят от температуры. В термодинамическом равновесии температуры всех частей тела одинаковы. Приводя его в тепловой контакт с исследуемым объектом и дождавшись установления равновесия между ними, можно определить температуру объекта. Таким прибором может быть газовый термометр.

Неоднозначность определения температуры даёт возможность использовать любую соответствующую исторически сложившейся традиции шкалу её измерения. В дальнейшем под термином температура будем подразумевать абсолютную температуру, измеренную в градусах Кельвина. Несмотря на теснейшую связь энергии и температуры, понятия эти существенно различны. Энергия – величина аддитивная, а температура – транзитивная.

Именно температуру, а не внутреннюю энергию, удобно использовать в качестве одной из основных независимых переменных в термодинамике. Энергия (если не говорить о теории относительности) вообще не имеет однозначного определения и, как и энтропия, не принадлежит к числу непосредственно измеряемых величин. Тем или иным способом может быть измерено лишь изменение энергии. В отличие от энтропии энергией обладает каждая молекула, тогда как энтропия, как и температура – характеристика макроскопического объекта.

Достаточно малые (но макроскопические) тела достигают равновесия сравнительно быстро. Это даёт возможность говорить о температурах частей тела, далёкого от равновесия, находящихся в состоянии локального равновесия. При фиксированном внешнем параметре температуры всех частей тела, вообще говоря, изменяются в ходе релаксации. Принято считать, что в равновесии энтропия достигает максимума. Но для такого утверждения нет оснований. Энтропия определена для равновесной системы. Для того, чтобы функция имела максимум в какой-либо точке, она должна быть определена в её окрестности. Свойства целого не определяются какой-либо совокупностью свойств его частей, хотя и зависят от них.

## МНОГОПАРАМЕТРИЧЕСКИЕ СИСТЕМЫ

Статистический подход применим как к изучению равновесных состояний макроскопических объектов и квазистатических процессов, так и к анализу любых состояний и процессов. Если говорить о газах, предметом теории в этом случае служит функция распределения числа молекул в объёме и в пространстве импульсов. Изменение функции распределения во времени определяется кинетическим уравнением. В состоянии равновесия, несмотря на непрекращающееся движение молекул, функция распределения не меняется. Функция распределения и вероятность нахождения молекулы в элементе фазового объёма совпадают (точнее – пропорциональны). В любых других состояниях системы функция распределения ничего общего с понятиями теории вероятностей не имеет.

Важнейшим понятием теории вероятностей является энтропия, определяемая средним значением логарифма вероятности [4]. Это наталкивает на возможность определения энтропии с помощью среднего значения логарифма функции распределения неравновесной, но локально равновесной системы. По мере приближения к равновесию определённая таким образом энтропия при условии постоянства внутренней энергии действительно растёт *(H-* теорема Больцмана). Но вероятности, относящиеся к равновесной системе, и вероятности, относящиеся к частям неравновесной системы, не совпадают. Они имеют не априорный, а апостериорный характер, и зависят от значений макроскопических параметров частей системы, связанных с параметрами других её частей. Взаимодействие макроскопических частей неравновесной системы определяется в случае газов не законами теории вероятностей, а законами гидродинамики. Разбиение на части по умолчанию обычно предполагает отнесение к части некоторого количества молекул. Но молекулы не являются носителями каких-либо термодинамических характеристик. Без детального определения способа разделения системы на части говорить о частях системы бессмысленно.

Поясним это на простых примерах. Рассмотрим горизонтально расположенный цилиндр с тремя поршнями: правый и левый – рабочие, а между ними – разделяющий. Обе части объёма между рабочими поршнями заполнены газами. Газы могут быть как одинаковыми, так и разными. Если разделяющий поршень неподвижный и теплоизолирующий, газы в обеих частях цилиндра образуют две совершенно независимые термодинамические системы. Но если разделяющий поршень неподвижный и теплопроводящий, или подвижный и теплоизолирующий, оба газа образуют единую систему. В первом случае в равновесии температуры обоих газов одинаковы, а во втором – одинаковы их давления. Соответственно, в обеих частях могут быть различными давления и температуры. В случае подвижного теплопроводящего разделяющего поршня и температуры и давления в обеих частях совпадают.

В случае отсутствия разделяющего поршня система является однопараметрической. В остальных упомянутых случаях условия равновесия соответствуют, вообще говоря, различным значениям транзитивных переменных. Система становится двухпараметрической. Внешними параметрами являются положения левого и правого рабочих поршней. В случае неподвижного теплопроводящего разделяющего поршня температуры обеих частей системы одинаковы, а давления в них могут быть различными. В случае подвижного теплоизолирующего – одинаковы давления, но температуры могут быть различными. В качестве термодинамических переменных в первом случае удобно принять общую температуру и оба объёма, а во втором – общее давление и обе температуры. Нетрудно убедиться, что ввиду независимости теплоёмкости газа от объёма в первом случае интегрирующий множитель пфаффовой формы всегда существует, а во втором он существует только при равенстве теплоёмкостей обоих газов. Нет интегрирующего множителя – нет и энтропии. Попытка "доопределить" энтропию системы как сумму энтропий составляющих её частей [6] смысла не имеет.

Термодинамическая энтропия и энтропия в теории вероятностей – существенно разные понятия. В этой книге, если это особо не оговорено, термин "энтропия" подразумевает термодинамическую энтропию, определённую с известной степенью произвола. Что же касается энтропии в теории вероятностей, она определена строго и однозначно, характеризуя любые ситуации, подчиняющиеся законам теории вероятностей.

Таким образом, добавление лишь одного параметра даёт возможность создать систему, не имеющую в состоянии равновесия единой температуры и поэтому не обладающую энтропией. Может сложиться впечатление, что отсутствие энтропии обусловлено термической неоднородностью. Но и в термически однородных системах может не быть интегрирующего делителя пфаффовой формы, а следовательно и энтропии. Температура системы, тем не менее, будучи измеряемой величиной, заведомо существует независимо от того, имеет ли пфаффова форма интегрирующий делитель или нет.

Системы с произвольным числом внешних параметров имеют пфаффову форму:

$$\delta Q = dE + \sum_{i=1}^{k} A_i(a_1, a_2, \ldots a_k, E) da_i \; , \tag{6}$$

где $\delta Q$ - переданное системе количество тепла, $dE$ - изменение её внутренней энергии, $da_i$ - изменения внешних параметров системы, $A_i$ - соответствующие им обобщённые силы, $k$ - число независимых внешних параметров [1].

В общем случае многопараметрических термически однородных термодинамических систем интегрирующий делитель пфаффовой формы (6) вообще не существует. При $k>1$ форма (6) имеет интегрирующий делитель лишь при выполнении условий Фробениуса [4], накладывающих весьма жёсткие ограничения на функции $A_i$. Для произвольной пфаффовой формы:

$$\omega_n = \sum_{i}^{n} y_i(x_1, x_2, \ldots x_n) dx_i \; , \tag{7}$$

$n$ – число независимых переменных, условия Фробениуса имеют вид [4,6]:

$$y_\alpha \left( \frac{\partial y_\gamma}{\partial x_\beta} - \frac{\partial y_\beta}{\partial x_\gamma} \right) + y_\beta \left( \frac{\partial y_\alpha}{\partial x_\gamma} - \frac{\partial y_\gamma}{\partial x_\alpha} \right) + y_\gamma \left( \frac{\partial y_\beta}{\partial x_\alpha} - \frac{\partial y_\alpha}{\partial x_\beta} \right) = 0 \tag{8}$$

$\alpha, \beta, \gamma = 1, 2, \ldots n \qquad \alpha \neq \beta \neq \gamma \neq \alpha$

По отношению к коэффициентам $y_i$ пфаффовой формы (7) условия Фробениуса есть совокупность $n!/[(n-3)!3!] = [n(n-1)(n-2)]/6$ условий, накладываемых на каждую тройку коэффициентов. Сама возможность удовлетворить набором $n$ функций значительно большему, чем n, количеству условий не исключена, но более чем проблематична.

## ВТОРОЕ НАЧАЛО

Существуют более дюжины различных эквивалентных формулировок второго начала термодинамики. В качестве аргумента в его пользу выдвигается даже решение, принятое Французской академией в XVIII веке задолго до возникновения термодинамики.

При этом нередко ссылаются на экспериментальные данные. Высказывания Р.Клаузиуса и В.Томсона по поводу второго начала термодинамики [7] только на первый взгляд соответствуют данным опыта, а в действительности имеют в своей основе лишь глубокое убеждение в невозможности осуществления вечного двигателя второго рода, сколько-нибудь серьёзных попыток создания которого и не могло быть тогда предпринято. В этой связи лишено какого бы то ни было основания утверждение Э.Ферми [8]: "Экспериментальное доказательство справедливости второго начала состоит главным образом в неудаче всех попыток сконструировать perpetuum mobile второго рода". Даже если бы такие попытки и предпринимались (о чем у нас нет определенных сведений), их неудача не может быть доказательством столь далеко идущего утверждения.

Введённое Р.Клаузиусом понятие энтропии, неразрывно связанное со вторым началом, лишь по звучанию напоминает энтропию в теории вероятностей. Оно не обладает той степенью общности, на какую претендует термодинамика в описании явлений природы. По этому поводу К.Трусделл писал [9]:
"Семь раз за последние тридцать лет я старался проследить аргументацию Клаузиуса,

пытавшегося доказать, что интегрирующий множитель существует в *общем случае,* и есть функция только температуры, одинаковая для всех тел, и семь раз это совершенно обескураживало меня... Я не могу даже объяснить, чего именно я не понимаю."

Сама манера изложения теории тепловых процессов, берущая начало в трудах основоположников термодинамики и сохранившаяся до наших дней, дала основание К.Трусделлу назвать термодинамику затхлым болотом обскурантизма [9]. В своей Нобелевской лекции И.Пригожин, отнюдь не подвергая сомнению начала термодинамики по существу, подчёркивал, что "через 150 лет после того, как второе начало было сформулировано, оно всё ещё представляет собой скорее программу, чем чётко очерченную теорию в обычном смысле этого понятия" [10].

Получившая широкое распространение трактовка начал термодинамики как обобщения опытных данных лишена серьёзных оснований. Достаточно упомянуть, что в середине XIX века относительная точность определения механического эквивалента тепла составляла десятки процентов [11]. В этих условиях говорить даже о первом начале как об опытном факте было по меньшей мере преждевременно. По существу, однако, первое начало – прямое следствие независимости физических законов от времени. Поэтому степень точности определения механического эквивалента тепла роли не играет.

С развитием статистической физики, все более стало утверждаться убеждение в статистической природе второго начала. Но далеко не все процессы в природе имеют вероятностный характер. Несмотря на то, что законы теории вероятностей накладывают свой отпечаток на многие явления природы, даже в рамках тепловых процессов он не является неизгладимым. Хотя хаотизация – характерный признак поведения замкнутых стохастических систем, близких к состоянию статистического равновесия, самоорганизация в природе не менее естественна, чем хаотизация. Так что второе начало, в отличие от первого, ни из каких общих принципов не вытекает.

Как первое, так и второе начало принято считать не просто началами термодинамики, но законами природы. Что касается первого начала, тут никакого сомнения не возникает. Совершенно иначе обстоит дело со вторым началом. Даже если допустить справедливость обоих начал на множестве всех квазистатических процессов, никаких оснований для распространения второго начала на процессы с интенсивными материальными и тепловыми потоками нет. Анализ всего множества неравновесных термодинамических процессов на предмет справедливости второго начала вряд ли когда-нибудь окажется осуществимым.

Принципиальные возражения по поводу второго начала были высказаны Дж.Максвеллом ("демон Максвелла"). Но они имели весьма экзотический характер. По сути дела второе начало утверждает невозможность создания термодинамического процесса преобразования тепла в работу без использования холодильника.

Непоколебимая вера во второе начало термодинамики тем более удивительна, что его опровержение буквально лежало на поверхности. Уже в середине XIX века необходимые для этого данные имелись. Одна из эквивалентных формулировок второго начала сводится к тому, что охлаждение термоизолированной системы в результате циклического изменения её внешних параметров невозможно. Для любой однопараметрической системы это действительно так. В случае газов адиабата на *(PV)* - диаграмме тем круче, чем меньше удельная теплоёмкость газа. При расширении газа в адиабатическом режиме производится работа, и газ охлаждается. Обратный процесс приводит газ в исходное состояние. Но если бы была возможность увеличить удельную теплоёмкость расширившегося газа, его сжатие пошло бы по более пологой адиабате, чем расширение. В результате возвращения объёма газа к исходному на сжатие затрачивалась бы меньшая работа, чем получалась при расширении, и исходная температура не была бы достигнута. Средств для изменения теплоёмкости газа при постоянном объёме не существует, но если привести его в тепловой контакт с другим объектом (жидкостью, твёрдым телом), поведение газа при изменении его объёма будет определяться суммарной теплоёмкостью. Адиабата на *(PV)* - диаграмме газа, контактирующего с другим объектом, будет более пологой, чем газовая. Увеличение теплоёмкости контактирующего объекта сделает её ещё более пологой. В результате расширение газа при минимальной теплоёмкости контактирующего объекта и его сжатие к исходному объёму при максимальной приведёт к охлаждению системы.

Разумеется, возникает вопрос: каким образом изменять теплоёмкость контактирующего объекта? Возможно ли это вообще, а если возможно, не будет ли это связано с затратой работы? Не превысит ли затрата работы на изменение теплоёмкости произведённой газом работы?

В качестве примера тела с регулируемой теплоёмкостью рассмотрим тонкий длинный стержень с положительным коэффициентом температурного удлинения. Энергия стержня есть сумма его внутренней энергии и потенциальной энергии, определяемой положением центра тяжести в гравитационном поле Земли. При горизонтальном положении стержня высота центра тяжести стержня в результате изменении его длины вследствие изменения температуры не меняется, а при вертикальном направление изменения высоты центра тяжести зависит от того, какая точка стержня закреплена. При закреплении нижней точки (стержень стоит) увеличение длины стержня приводит к подъёму центра тяжести, а при закреплении верхней точки (стержень висит) – к его опусканию. Соответственно, и теплоёмкость в первом случае выше теплоёмкости во втором.

Процесс охлаждения термоизолированной системы может быть следующим. В исходном состоянии температура равна температуре термостата $T_0$. Стержень стоит в вертикальном положении. Объём газа минимален. На первой стадии стержень переводится в висячее положение, поворачиваясь вокруг точки опоры. При этом производится некоторая работа. Температура системы остаётся неизменной. На второй стадии газ расширяется с производством работы. Температура системы понижается, длина стержня сокращается. Третья стадия – возврат стержня в вертикальное положение. В результате центр тяжести стержня будет ниже исходного. Затрачиваемая на это работа меньше работы, полученной на первой стадии. Температура остаётся неизменной. На заключительной стадии объём возвращается к исходному, и происходит нагрев системы. Поскольку сжатие газа происходит при большей теплоёмкости системы, чем расширение, работа, затраченная на сжатие, заведомо меньше работы, произведённой при расширении. Таким образом, как газ, так и стержень произвели работу. Это может произойти только в результате понижения температуры системы. При этом длина стержня не вернулась к исходной. Возврат системы в исходное положение возможен в результате установления теплового контакта с термостатом.

Адиабата газа, находящегося в тепловом контакте со стержнем, определяется уравнением:
$$(C + C_V)dT + PdV = 0, \qquad (9)$$

где *C* - теплоёмкость стержня. На второй стадии рассмотренного выше процесса $C = C_{down}$, а на четвёртой - $C = C_{up}$. В результате заключительная стадия приводит к температуре:
$$T_{fin} = T_0 \left(\frac{V_0}{V_{max}}\right)^{\frac{N(C_{up} - C_{down})}{(C_V + C_{up})(C_V + C_{down})}}, \qquad (10)$$

где $V_0$ - исходный объём газа, $V_{max}$ - его максимальный объём. Поскольку показатель степени в (10) положителен, $T_{fin}$ заведомо меньше $T_0$. Не прибегая к справочникам теплофизических параметров веществ, очевидно, что в реальных условиях даже при очень больших расширениях газа эффект охлаждения исключительно мал, но его принципиальное значение невозможно переоценить. Приведённый пример – хоть и простое, но вряд ли более убедительное возражение второму началу, нежели демон Максвелла. Кстати, недавно в Интернете появилось сообщение о создании в Японии демона Максвелла [12].

## ПРЕОБРАЗОВАНИЕ ТЕПЛА В РАБОТУ В ТЕРМИЧЕСКИ ОДНОРОДНОЙ ДВУХПАРАМЕТРИЧЕСКОЙ СИСТЕМЕ

Изменение термодинамических характеристик системы в ходе преобразования тепла в работу – трудноосуществимый процесс. Зависимость термодинамических параметров от условий проведения процесса в случае твёрдых тел и жидкостей пренебрежимо мала. Теплоёмкости $C_{up}$ и $C_{down}$ различаются хорошо, если в пятом-шестом знаке. Куда более существенно зависят от условий проведения процесса теплоёмкости газов.

Исследуем двухпараметрическую термически однородную систему газов, разделённых неподвижной теплопроводной перегородкой. Независимыми переменными такой системы служат оба объёма газов и их общая температура. Пфаффова форма такой системы, с учётом уравнения состояния идеального газа, в общепринятых обозначениях имеет вид:

$$\delta Q = CdT + \frac{N_1 T}{V_1} dV_1 + \frac{N_2 T}{V_2} dV_2, \qquad (11)$$

где С – теплоёмкость системы, равная сумме теплоёмкостей обоих газов. Поскольку теплоёмкость газа может зависеть только от температуры, форма (11) заведомо интегрируема, и, казалось бы, никаких осложнений со вторым началом ожидать не приходится. Однако, такое заключение было бы слишком поспешным. Дело в том, что теплоёмкость зависит не только от свойств вещества, вступающего в теплообмен, но и от условий, в которых этот теплообмен происходит. Основная идея заключается в возможности управления термодинамическими характеристиками системы в ходе теплового процесса [13].

Рассмотрим теплоизолированную систему, состоящую из двух объёмов, заполненных газами, разделённых неподвижной теплопроводной перегородкой. Тепловой контакт обеспечивает равенство температур обоих газов в ходе квазистатического процесса. Справа от перегородки расположен горизонтально цилиндр с рабочим поршнем, а слева – два связанных между собой цилиндра: один – горизонтальный с рабочим поршнем, другой – вертикальный с тяжёлым поршнем (Рис.1а). Оба рабочих поршня являются внешними параметрами системы.

При лежащем на дне вертикального цилиндра тяжёлом поршне максимально допустимое давление газа в левой части системы определяется весом тяжёлого поршня. При более низких давлениях объём газа в левой части однозначно определяется положением левого рабочего поршня. В этих условиях движение левого рабочего поршня приводит к изменению объёма и давления газа в объёме $V_1$, а пфаффова форма совпадает с пфаффовой формой двух газов, разделённых теплопроводной перегородкой. Теплоёмкости обоих газов при этом равны их теплоёмкостям при постоянном объёме.

Если тяжёлый поршень не лежит на дне вертикального цилиндра, давление газа в левой части постоянно, и определяется весом поршня. При движении левого рабочего поршня тяжёлый поршень тоже перемещается, причём объём $V_1$, температура и давление газа в этом объёме остаются неизменными. Левая часть системы потенциальна [14]. Теплоёмкость газа в левой части равна его теплоёмкости при постоянном давлении.

И в том, и в другом случае пфаффова форма системы имеет интегрирующий множитель. Но если в ходе термодинамического процесса происходит опускание тяжёлого поршня на дно, а затем и возврат на высоту свободного движения, говорить о единой для всего процесса пфаффовой форме, тем более о её интегрирующем множителе, не приходится. Ни теплоёмкость системы, ни её пфаффова форма никакой гладкой функцией не описываются. Тем не менее, такие процессы не представляет особого труда изучить в общем виде.

Возможен следующий четырёхстадийный процесс. На первой стадии процесса рабочий поршень в левом горизонтальном цилиндре движется налево вплоть до упора тяжёлого поршня в дно вертикального цилиндра (Рис.1b). В результате производится работа, равная уменьшению потенциальной энергии тяжёлого поршня в поле силы тяжести, а левая часть системы лишается свойства потенциальности.

На второй стадии увеличивается объём правой части системы в результате движения правого поршня направо с производством работы и охлаждением системы в целом (Рис.1c). Давление газа в левой части уменьшается. Охлаждение производится настолько, чтобы при возвращении левого рабочего поршня в исходное положение давление газа в левой части не достигало исходного. При достаточно глубоком охлаждении энергия газа в левой части практически полностью переходит в правую часть, а затем — в работу. В пределе неограниченного расширения газа в правой части внутренняя энергия газа обеих частей системы полностью переходит в работу.

Далее рабочий поршень в левой части возвращается в исходное положение (Рис.1d). Затраченная на возврат рабочего поршня работа заведомо меньше работы, произведённой на первом этапе. Нагревом системы в целом при этом можно пренебречь. На завершающей стадии возвращается к исходному положению правый рабочий поршень, и нагревается система в целом (Рис.1e). В начале стадии сжатия структура системы отличается от структуры на стадии расширения объёмом газа в левой части. Поскольку теплоёмкость газа не зависит от объёма, тепловые характеристики системы при сжатии тождественны характеристикам при расширении. Нагрев газа в левой части идёт при постоянном объёме, и правая часть проходит через ту же последовательность состояний, что и на второй стадии, в обратном порядке. По достижении исходного давления газа в левой части в результате нагрева эта часть системы вновь становится потенциальной, и её тепловые параметры кардинально меняются. Дальнейший нагрев системы производится не при постоянном объёме газа в левой части, а при постоянном давлении. Рост давления и температуры, а также затрачиваемой на сжатие работы, по мере уменьшения объёма правой части замедляется. В результате возврата к исходному положению как левого (на третьей стадии), так и правого (на заключительной стадии) поршней исходная температура не достигается.

Таким образом, возврат к исходным значениям внешних параметров системы приводит к её охлаждению и производству работы, что противоречит одной из множества эквивалентных формулировок второго начала термодинамики. Для восстановления всех исходных термодинамических характеристик системы необходимо лишь установить тепловой контакт с имеющим исходную температуру термостатом.

В качестве конкретного примера рассмотрим случай идеального одноатомного газа.

Числа молекул газов в обеих частях системы $N$ и их исходные давления $P_0$ и объёмы $V_0$ будем считать одинаковыми. Изменения параметров системы при изменении лишь одного внешнего параметра определяются известными формулами для адиабатического процесса [5]:

$$T^\gamma P^{1-\gamma} = const, \qquad TV^{\gamma-1} = const, \qquad PV^\gamma = const. \qquad (12)$$

Теплообмен между обеими частями системы приводит к увеличению теплоёмкости каждой части. В случае постоянства объёма контактирующей части происходит удвоение теплоёмкости изменяющейся части, а в случае постоянства давления к теплоёмкости изменяющейся части добавляется теплоёмкость контактирующей части при постоянном давлении. Соответственно показатель "адиабаты" для первого случая будет $\gamma_1 = 1 + 1/2c_v$, а для второго - $\gamma_2 = 1 + 1/(2c_v + 1)$, где $c_V$ - теплоёмкость газа при постоянном объёме в расчёте на одну молекулу. Для одноатомного газа $c_V = 3/2$, так что $\gamma_1 = 4/3$, а $\gamma_2 = 5/4$.

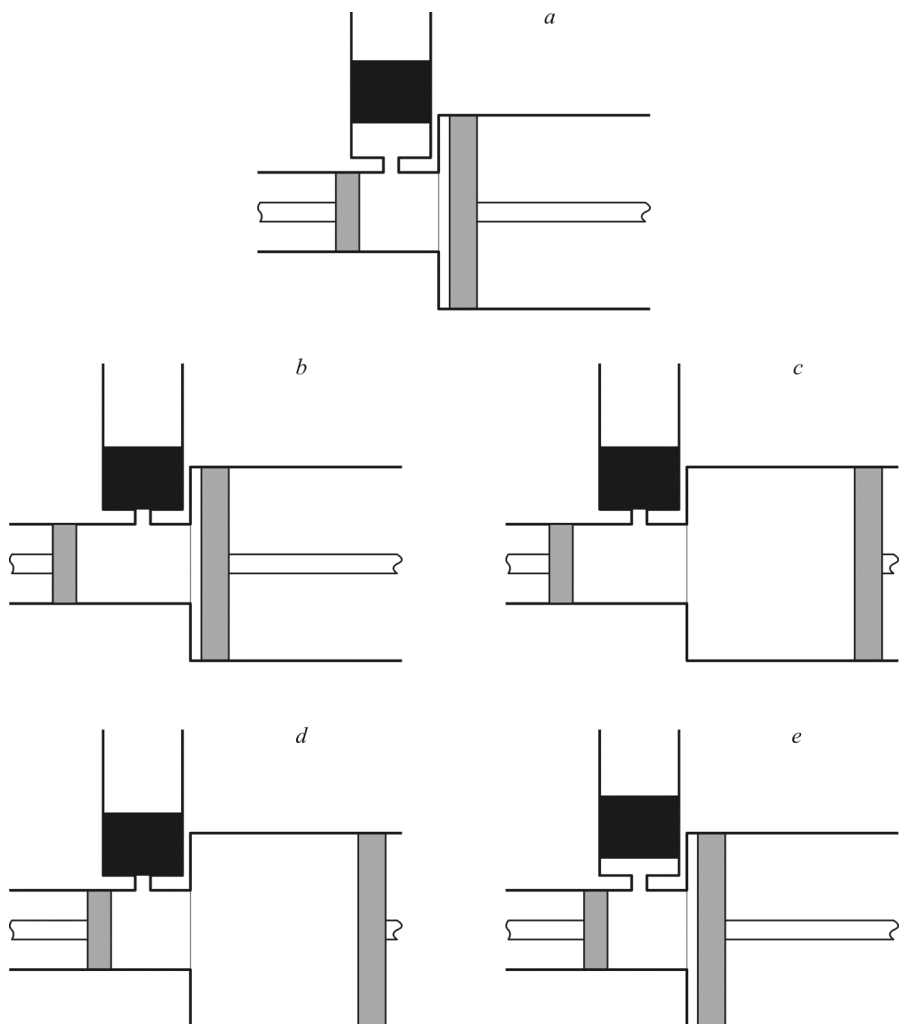

Рис. 1. Состояние системы на различных этапах процесса
a) исходное положение   b) конец первого этапа   c) конец второго этапа
d) конец третьего этапа   e) конец заключительного этапа

Исследуем вначале случай постоянства показателя "адиабаты" на каждом этапе циклического процесса. В ходе первого этапа левый рабочий поршень движется налево вплоть до упора тяжелого поршня в дно вертикального цилиндра. Тепловые характеристики системы не меняются.

Увеличение объёма газа правой части системы на втором этапе и её охлаждение определяются из условия возврата давления газа в левой части к исходному значению в конце третьего этапа (при возвращении левого рабочего поршня в исходное положение). Изменение показателя "адиабаты" происходит при переходе от третьего этапа к заключительному. Очевидно, температура системы в конце третьего этапа $T_3$ в $V_0/V_3$ раз меньше исходной, где $V_3$ - объём газа в левой части в конце третьего этапа. Температура системы в начале третьего этапа (или в конце второго) $T_2$ определяется из условия: $T_2V_0^{1/3} = T_3V_3^{1/3}$. Так что $T_2 = (V_3/V_0)^{4/3}T_0$. Объём правой части в конце второго этапа $V_2$ определяется аналогичным условием. В результате $V_2 = (V_0/V_3)^4 V_0$. Итоговая температура $T_f$ определяется условием: $T_f V_0^{1/4} = T_3 V_2^{1/4}$. В итоге $T_f = T_0$. Глубина охлаждения системы определяется величиной $V_2$. При $V_2 < V_0(V_0/V_3)^4$ переход от режима постоянного объёма к режиму постоянного давления газа в левой части происходит уже в ходе третьего этапа процесса. При $V_2 > V_0(V_0/V_3)^4$ этот переход осуществляется на заключительном этапе.

Если $V_2 < V_0(V_0/V_3)^4$, ввиду недостаточного охлаждения системы на втором этапе, при сжатии газа в левой части достигается исходное давление, и система вновь обретает потенциальную часть. Дальнейшее движение левого рабочего поршня приводит лишь к подъёму тяжёлого поршня и не меняет тепловых параметров системы. Температура при переходе к постоянному давлению газа в левой части $T_c$ определяется из условия: $T_c P_0^{-1/4} = T_2 P_2^{-1/4}$, где $T_2$ - температура системы в конце второго этапа, а $P_2$ - давление газа в левой части в конце второго этапа: $P_2 = NT_2/V_0$. Поскольку $P_0 = NT_0/V_0$, $T_c = T_2(T_0/T_2)^{1/4}$. Температура системы при переходе к заключительному этапу $T_4 = T_c$. Итоговая температура определяется условием: $T_f V_0^{1/4} = T_4 V_2^{1/4}$. В результате $T_f = T_0$. Так что при постоянстве показателя "адиабаты" на заключительном этапе никакого охлаждения не происходит.

Ситуация кардинально меняется при $V_2 > V_0(V_0/V_3)^4$. В этом случае возврат рабочего поршня в исходное положение и переход к заключительному этапу происходят при давлении газа в левой части меньшем исходного. Температура перехода к постоянному давлению газа в левой части $T_c$ определяется из уравнения состояния идеального газа: $T_c = P_0 V_3/N = T_0 V_3/V_0$, а объём газа в правой части системы $V_c$ – из условия: $T_c V_c^{1/3} = T_3 V_2^{1/3}$. Итоговая температура системы определяется условием: $T_f V_0^{1/4} = T_c V_c^{1/4}$. В результате $T_f = (V_3/V_0)^{3/4} T_0$.

Характерно, что итоговая температура не является аналитической функцией параметров системы. При недостаточно глубоком охлаждении на втором этапе процесса итоговая температура возвращается к исходной независимо от глубины охлаждения. Достаточно глубокое охлаждение уменьшает итоговую температуру не при неограниченном расширении газа на втором этапе, когда оно очевидно, а сразу же по выполнении условия $V_2 > V_0(V_0/V_3)^4$, Дальнейшее охлаждение системы на втором этапе итоговую температуру не меняет. Такого рода "бистабильность" не состояния, а результата процесса, ранее в термодинамике не встречалась.

На Рис.1 отображена последовательность состояний системы в процессе при $V_2 > V_0(V_0/V_3)^4$, а на Рис.2 и Рис.3 представлен вид диаграмм давление – смещение рабочего поршня при различных значениях параметров системы. Положение правого поршня однозначно определяет объём газа в правой части системы. Для левой части системы такое взаимно однозначное соответствие существует лишь при давлениях меньших исходного.

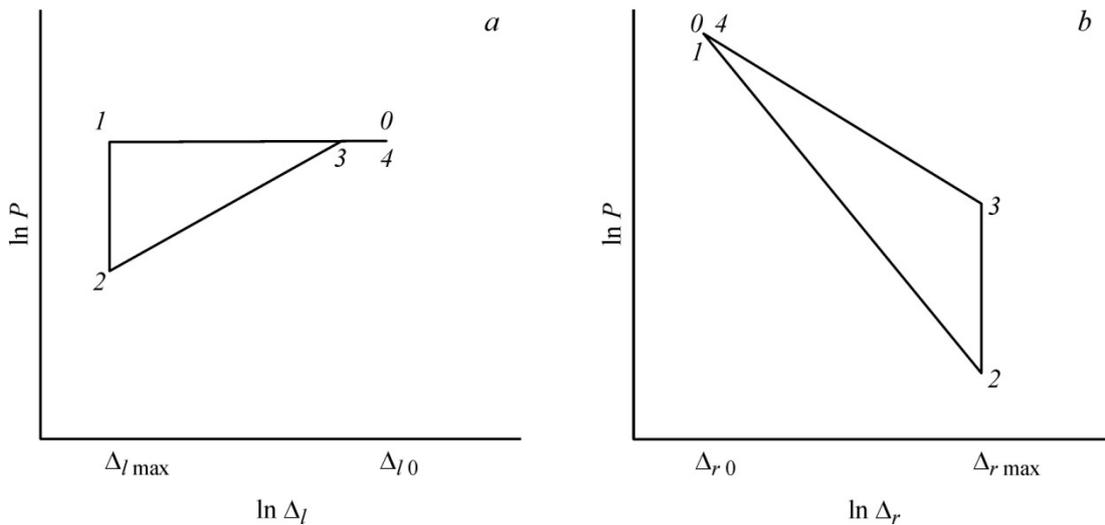

Рис. 2 Вид диаграммы давление – смещение рабочего поршня при $V_2 < V_0(V_0/V_3)^4$

a) Левый поршень  b) Правый поршень

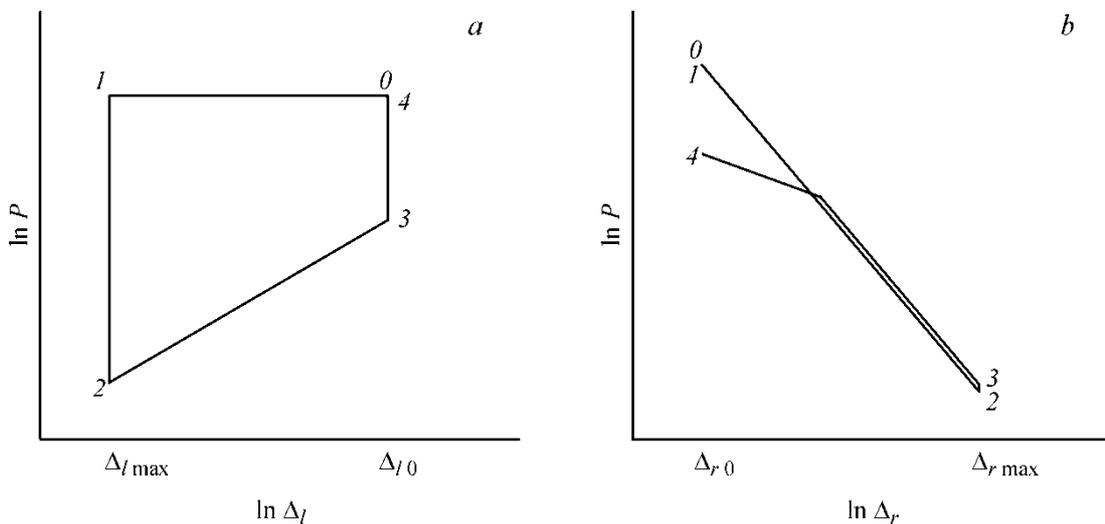

Рис.3 Вид диаграммы давление – смещение рабочего поршня при $V_2 > V_0(V_0/V_3)^4$

a) Левый поршень  b) Правый поршень

Термодинамические кривые (в логарифмическом масштабе – прямые) определяют величину произведённой работы. Работа за цикл, произведённая левым поршнем, всегда положительна. При $V_2 < V_0(V_0/V_3)^4$ работа за цикл, произведённая правым поршнем, отрицательна и по абсолютной величине равна работе левого поршня. В предельном случае $V_2 \to \infty$ работа за цикл, произведённая правым поршнем, положительна. При $V_2 < V_0(V_0/V_3)^4$ цикл замкнут по всем переменным, а при $V_2 > V_0(V_0/V_3)^4$ - только по внешним параметрам.

# ЗАКЛЮЧЕНИЕ

В изложенных выше примерах полного преобразования тепла в работу наряду с законом сохранения энергии по умолчанию применялось бесспорное утверждение: "при наличии теплового контакта между телами тепло само переходит от более нагретого тела к менее нагретому", которое отнюдь не эквивалентно второму началу. Восстановление исходного состояния системы путём нагрева необратимо. Поэтому, строго говоря, все рассмотренные процессы не укладываются в рамки термостатики. Но этот выход столь прост и нагляден, что вряд ли смогут возникнуть сомнения в его правомерности.

Энтропия характеризует степень упорядоченности стохастической системы, подчиняющейся законам теории вероятностей. Теория вероятностей распространима лишь на крайне ограниченный круг стохастических систем и процессов. Действительно, даже в простейших случаях вероятность определяется как предел отношения числа реализаций события к общему числу попыток при неограниченном увеличении числа попыток. В общем случае такого предела может и не существовать. Статистические системы или процессы в общем случае характеризуются хаотическим поведением. Надёжная база для их исследования отсутствует. В «Математической энциклопедии» [4] нет даже определения понятия «хаос». Применение теории вероятностей возможно лишь в том случае, если состояния системы обладают определёнными вероятностями. В однопараметрических системах это условие соблюдается всегда. Изменение внешнего параметра или температуры приводит к изменению не только вероятностей, но и выборочного пространства. С этом смысле вероятности не априорны, а апостериорны (условны). Мерой упорядоченности системы служит энтропия. В равновесии термодинамическая энтропия и энтропия в теории вероятностей по существу совпадают. В многопараметрических системах изменение любого внешнего параметра приводит к изменению не только характеристик соответствующей части системы, но и всех её частей. Многопараметрические системы в состоянии равновесия могут быть и термически неоднородными. Такие системы, вообще говоря, неинтегрируемы, а потому и энтропии иметь не могут. Но и термически однородные системы в случае зависимости теплоёмкости от давления неинтегрируемы и, как мы убедились, допускают полное преобразование тепла в работу.

Даже в рамках квазистатических процессов макроскопические тела могут энтропии и не иметь как изначально, в случае термически неоднородных тел, так и по отношению к процессу, в ходе которого изменяются термодинамические параметры системы. Именно второе начало и принцип максимума энтропии обеспечивают устойчивость локального термодинамического равновесия и равенство температуры всех частей тела в равновесии. При разделении тела на части независимо от того, равновесно оно или нет, каждая часть представляет собой однопараметрическую систему, а по отношению к ней вся аргументация, опирающаяся на второе начало, сохраняет свою силу. В условиях локального равновесия все части системы обладают и температурой и энтропией. В этих случаях второе начало само является следствием интегрируемости соответствующей пфаффовой формы. Но система в целом ни температуры ни энтропии не имеет. Хотя термодинамическая энтропия и энтропия в теории вероятностей имеют разные определения, термодинамическая энтропия имеет смысл лишь в применении к равновесным системам, когда функция распределения и распределение вероятностей по существу совпадают.

Для выравнивания температур частей тела при отсутствии теплонепроницаемых перегородок достаточно приведённого выше утверждения, заключённого в кавычки, или его эквивалента. Одного только первого начала недостаточно как для определения температуры, ввиду его неоднозначности, так и для создания возможности её измерения. Каким именно должно быть «второе начало» – вопрос открытый, но во всяком случае не таким, каким представлял его А.Эйнштейн. Возможным вариантом может быть: «Все части равновесной системы, находящиеся в тепловом контакте, имеют одинаковую температуру».

В термодинамике однопараметрических систем единственным способом уменьшения энтропии системы служит её охлаждение. Как температура, так и энтропия – величины, обладающие тем свойством, что в теплоизолированной системе в результате любых циклических изменений внешних параметров, независимо от того, обратимы они или нет, они могут, в соответствии с существующими представлениями, только возрастать. Однако снижение температуры многопараметрической системы возможно не только при охлаждении, но и в результате циклического изменения внешних параметров. Снижение температуры служит, пожалуй, более универсальным свидетельством «наведения порядка» в системе в целом, нежели снижение энтропии, поскольку эта энтропия в общем случае не существует. Но даже в тех случаях, когда термодинамическая энтропия существует, она определена с точностью до произвольной постоянной. «Третье начало», в отличие от нулевого, к основным принципам отнесено быть не может.

Охлаждение теплоизолированной термодинамической системы в результате циклического изменения её внешних параметров можно рассматривать как процесс самоорганизации. Хотя общих критериев степени хаотичности системы и не существует, охлаждение всех её частей заведомо свидетельствует о происшедшей в ней самоорганизации [15, 16, 17].

Самоорганизация – одно из неоспоримых, но и загадочных проявлений Природы. Жизнь немыслима без самоорганизации. Но и в неживой природе процессы самоорганизации дают о себе знать. Потрясающие воображение торнадо – наглядные тому примеры (если вырывание деревьев с корнем, разрушение домов и гибель людей можно рассматривать как результат самоорганизации). Общепринято, что самоорганизация возможна лишь в открытых системах (см. [15] и цитированную там литературу). Замкнутая многопараметрическая система есть совокупность статистических систем, причём каждая её часть обладает своим выборочным пространством. Отдельные части системы можно рассматривать как открытые по отношению к остальным её частям. Даже в замкнутых двухпараметрических системах квазистатический процесс в определённых условиях сопровождается самоорганизацией. При этом внешние параметры играют роль управляющих.

Использование возможности управления термодинамическими характеристиками в целях повышения организованности термодинамической системы не исчерпывает способы такого повышения. Для этого могут применяться внешние поля, а также инерционные силы, возникающие в ходе макроскопического движения сред. В результате даже в условиях равновесия в сопутствующей системе координат могут возникать значительные градиенты давления и температуры газов. Температурная однородность характерна только в случае отсутствия силовых полей.

Принято считать, что равновесная функция распределения газа в поле силы тяжести определяется барометрической формулой. Температура газа от высоты не зависит. В то же время хорошо известно, что по мере увеличения высоты в атмосфере Земли падает не только давление, но и температура воздуха. Этот факт обычно объясняется неравновесностью атмосферы. Конвекция газа в поле силы тяжести вследствие зависимости давления газа от высоты и его малой теплопроводности может приводить к возникновению температурных градиентов. Этот процесс наряду с нагревом воздуха от поверхности Земли должен приводить к уменьшению температуры с высотой, ограниченному условием устойчивости механического равновесия (адиабатическое расслоение атмосферы). Градиент температуры может достигать величины порядка десяти градусов на километр [1].

В условиях термодинамического равновесия при отсутствии гравитационного поля внутренняя энергия, давление и температура газа от координат не зависят. Равновесны и все элементы массы газа. Каждый элемент обладает своими термодинамическими характеристиками. (Локальное термодинамическое равновесие.) Не зависят от координат не только давление и температура, но и удельная энтропия. Ввиду малой теплопроводности газа движение малых элементов массы можно считать адиабатическим. В равновесии перемешивание не меняет состояния системы, что возможно лишь при независимости удельной энтропии от координат. Как параметры системы в целом, так и параметры элементов массы удовлетворяют закону сохранения энергии, известным образом определяемому пфаффовой формой.

Температура есть интегрирующий делитель пфаффовой формы, и в силу этого является функцией его внутренней энергии и объёма. Но поскольку температура, как интегрирующий делитель пфаффовой формы, зависит от внутренней энергии, при условии зависимости внутренней энергии в поле силы тяжести от высоты в термодинамическом равновесии температура не может быть независимой от высоты.

В общем случае никакой зависимости между внутренней энергией и температурой нет. Энергия, в отличие от температуры, не является измеримой величиной и определена с точностью до произвольной постоянной. Изменение высоты расположения элемента массы газа в гравитационном поле при фиксированном объёме существенно меняет внутреннюю энергию газа, но никак не влияет на его температуру. И внутренняя энергия и объём — независимые переменные, полностью характеризующие состояние равновесной термодинамической системы. Тесная связь внутренней энергии и температуры элемента массы газа возникает лишь в случае изменения обеих переменных, определяющих его состояние.

Зная зависимость внутренней энергии от высоты, казалось бы нетрудно определить и зависимость температуры от высоты. Но для этого необходимо знать значение теплоёмкости газа, а его теплоёмкость зависит от условий нагрева или охлаждения. При наличии поля силы тяжести не только внутренняя энергия, но и давление зависят от высоты. Самодиффузия в газе выражается в перестановке различных элементов массы, при равновесии не меняющей его состояния. Перестановка элементов массы по высоте в этом случае связана с изменением в них давления, а следовательно их объёмов и температур.

Определение равновесной функции распределения из условия максимума энтропии — типичная изопериметрическая задача. Обычно предполагаемое постоянство неопределённых множителей Лагранжа [5] учитывает возможность обмена энергией между различными элементами массы, но не предусматривает возможность их перемещения, и по умолчанию приводит к независимости температуры от высоты. Это соответствует максимуму энтропии при условии не только сохранения внутренней энергии и числа частиц в системе, но и запрета свободного перемещения молекул. Снять этот запрет можно лишь допустив независимость от координат удельной энтропии. В этом случае весь объём системы доступен каждому элементу массы и каждой молекуле газа, причём перестановка различных элементов массы не меняет ни энергии, ни энтропии системы.

В общепринятых обозначениях равновесный градиент температуры определяется выражением:

$$\frac{dT}{dz} = \frac{\gamma - 1}{\gamma} \frac{T}{P} \frac{dP}{dz} = -\frac{\gamma - 1}{\gamma} \frac{mg}{k} \qquad (13)$$

Он совпадает с границей конвективной устойчивости [1]. Более быстрое падение температуры с высотой приведёт к механической неустойчивости системы и развитию конвекции, а более медленное — к снижению энтропии. Не только механическое, но и термодинамическое равновесие соответствуют максимально возможному падению температуры с высотой, допустимому условием механического равновесия, и максимальной энтропии. Только в состоянии безразличного механического равновесия и максимума энтропии перемешивание масс газа не меняет ни энергии, ни энтропии системы. Наличие градиентов давления и температуры в равновесном состоянии не приводит к возникновению материальных или тепловых потоков.

Линейное падение температуры с высотой не меняет существенным образом ни уравнение Эйлера, ни вид равновесной функции распределения вероятностей. Нужно лишь заменить $T$ на $T_0 - (z - z_0)(\gamma - 1)mg/\gamma k$, где $T_0$ - температура на уровне $z_0$. Зависимость температуры от высоты определяет при достаточно больших $z$ резкую границу газа, соответствующую обращению температуры в нуль. Но если при приближении к границе температура падает с высотой линейно, то давление падает экспоненциально с неограниченно увеличивающимся показателем экспоненты.

Обнаружить ничтожный температурный градиент в воздухе, обусловленный силой тяжести, в лабораторных условиях вряд ли удастся. Сила тяжести слишком слаба. Хотя основные понятия термодинамики определены для неподвижных тел и имеют смысл лишь в сопутствующей системе координат, термодинамическое равновесие по крайней мере на локальном уровне возможно и в случае движущихся тел, в частности — в центрифугах. Применяя современные центрифуги и тяжелые газы неоднородность температуры в равновесном состоянии наверняка можно заметить.

Уникальную возможность для исследования влияния ускорения на равновесную температуру газа предоставляют вихревые трубы. При радиусе трубы порядка одного сантиметра и инжекции газа с околозвуковой скоростью характерные ускорения в вихревом потоке на шесть порядков величины превосходят ускорение свободного падения. Обусловленные таким огромным ускорением градиенты температуры достигают десятка градусов не на километр, а на миллиметр. Соответственно разность температур "горячего" и "холодного" потоков газа из вихревой трубы доходит до многих десятков градусов [18].

Сама возможность влияния силы тяжести на температуру газа в условиях термодинамического равновесия находится в прямом противоречии с общепринятыми формулировками второго начала термодинамики [1]. Это утверждение бесспорно. Линейная зависимость температуры от высоты — ещё одно неоспоримое свидетельство необходимости пересмотра содержания формулировок второго начала.

Среди объектов научных интересов Я.И.Френкеля была и метеорология. Рассуждая о вопросах метеорологии, Яков Ильич любил повторять: "Облако – это не вещь. Облако – это процесс". По существу любой макроскопический объект неразрывно связан с тем или иным процессом в нём происходящим. Но в облаке это проявляется особенно ярко.

Творцы термодинамики и её адепты, не разобравшись толком в природе вещей, поспешили высказать категорическое суждение о возможных процессах с ними происходящих. Облачённое в наукообразную форму суждение было провозглашено законом природы [19]. По сути дела это был лишь реверанс в сторону решения Французской академии. В результате инженерная мысль и общественное мнение на многие годы были дезориентированы.

В рамках квазистатических процессов никаких ограничений на возможность полного преобразования тепла в работу не существует. Но практического интереса они не представляют, поскольку их мощность равна нулю. Реальные системы преобразования тепла в работу неразрывно связаны с движением рабочего вещества, поэтому исследование их свойств лежит за рамками существующей термодинамики.

# СПИСОК ЛИТЕРАТУРЫ